\documentclass[a4paper,12pt]{article}
\pdfoutput=1 

\usepackage{jheppub} 

\usepackage[T1]{fontenc} 
\usepackage{color,graphicx,epsfig}
\usepackage{ifpdf}
\usepackage{amsmath}
\usepackage{bm}
\usepackage[english]{babel}
\usepackage{graphicx}%
\usepackage{amsfonts}%
\usepackage{amssymb}
\usepackage{braket}
\usepackage{hyperref}
\usepackage{enumerate}
\usepackage{subfigure}

\usepackage[T1]{fontenc}
\usepackage{slashed}
\usepackage[utf8]{inputenc}
\usepackage{xspace}
\usepackage{ulem}
\usepackage{array}
\usepackage{ulem,fancyvrb}
\usepackage{xcolor}
\newcommand{\tabincell}[2]{\begin{tabular}{@{}#1@{}}#2\end{tabular}}
\title{\boldmath Detecting an axion-like particle with machine learning at the LHC}

\author[1]{Jie Ren,}
\author[2,3]{Daohan Wang,}
\author[4]{Lei Wu,}
\author[2,3]{Jin Min Yang,}
\author[5]{Mengchao Zhang}

\affiliation[1]{School of Aerospace Engineering, Beijing Institute of Technology, Beijing 100081, China}
\affiliation[2]{CAS Key Laboratory of Theoretical Physics, Institute of Theoretical Physics,
Chinese Academy of Sciences, Beijing 100190, China}
\affiliation[3]{School of Physical Sciences, University of Chinese Academy of Sciences,
Beijing 100049, China}
\affiliation[4]{Department of Physics and Institute of Theoretical Physics, Nanjing Normal University, Nanjing 210023, China}
\affiliation[5]{Department of Physics and Siyuan Laboratory, Jinan University, Guangzhou 510632,  China}

\emailAdd{jieren21@live.com}
\emailAdd{wangdaohan@mail.itp.ac.cn}
\emailAdd{leiwu@njnu.edu.cn}
\emailAdd{jmyang@itp.ac.cn}
\emailAdd{mczhang@jnu.edu.cn}

\abstract{Axion-like particles (ALPs) appear in various new physics models with spontaneous global symmetry breaking. When the ALP mass is in the range of MeV to GeV, the cosmology and astrophysics bounds are so far quite weak. In this work, we investigate such light ALPs through the ALP-strahlung production processes $pp \to W^\pm a, Z a$ with the sequential decay $a \to \gamma\gamma$ at the 14 TeV LHC with an integrated luminosity of 3000 fb$^{-1}$ (HL-LHC). Building on the concept of jet image which uses calorimeter towers as the pixels of the image and measures a jet as an image, we investigate the potential of machine learning techniques based on convolutional neural network (CNN) to identify the highly boosted ALPs which decay to a pair of highly collimated photons. With the CNN tagging algorithm, we demonstrate that our approach can extend current LHC sensitivity and probe the ALP mass range from 0.3~GeV to 5~GeV. The obtained bounds are stronger than the existing limits on the ALP-photon coupling.}

\begin{document} 
\maketitle
\flushbottom
\section{Introduction}

Many extensions of the Standard Model (SM) predict the existence of light pseudo-scalars, the so-called axion-like particles (ALPs). In general, they are predicted in any models with spontaneous breaking of a global $U(1)$ symmetry~\cite{Peccei:1977hh,Weinberg:1977ma,Wilczek:1977pj,Kim:1979if} and also appear in supersymmetry (SUSY) with dynamical SUSY breaking~\cite{Bagger:1994hh} or spontaneously  R-symmetry breaking~\cite{Bellazzini:2017neg} as well as in compactifications of string theory~\cite{Svrcek:2006yi,Arvanitaki:2009fg,Cicoli:2012sz}. Besides, they may play a crucial role in solving the hierarchy problem~\cite{Graham:2015cka} and be related to the electroweak phase transition~\cite{Ballesteros:2016euj}. So far such ALP particles have been being searched at the Large Hadron Collider (LHC) and their phenomenological studies are being extensively studied nowadays. 

\begin{figure}[ht]
\begin{center}
\includegraphics[width=15cm]{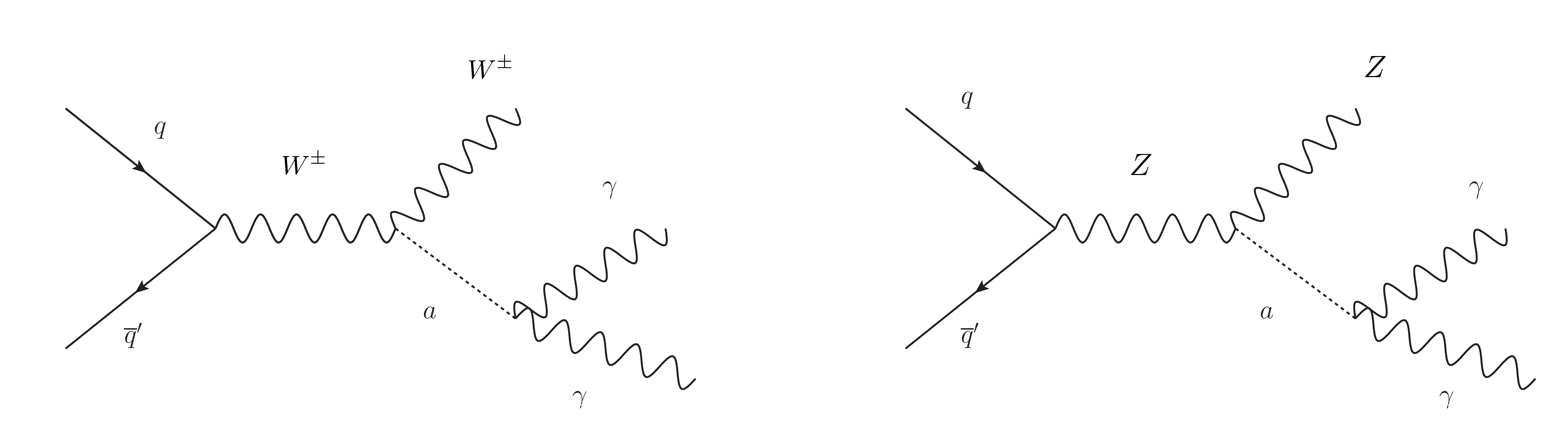}
\vspace{-0.5cm}
\caption{Feynman diagrams of ALP-strahlung production processes $pp \to W^\pm a$ and $pp \to Za$ with $(a\to \gamma\gamma)$ at the LHC}.
\label{Feynman_diagrams}
\end{center}
\end{figure}

Of course, for the phenomenological studies,  the ALP masses and couplings to SM particles are the most important parameters. A considerable region of these parameters has already been probed by the cosmological observations, the low-energy experiments and the high energy colliders~\cite{Dobrescu:2000jt,Chang:2006bw,Toro:2012sv,Draper:2012xt,Ellis:2012zp,Mimasu:2014nea,Knapen:2016moh,Barrie:2016ntq,Bauer:2017ris,Brivio:2017ije,Bauer:2018uxu,Ebadi:2019gij,Takahashi:2020bpq,Han:2020dwo,Gu:2021lni,Athron:2020maw}:
\begin{itemize}
\item[(i)] 
For an ALP wih mass $m_a\gtrsim 5$ GeV, the LEP and current LHC experiments have good sensitivities and a considerable region of parameter space has been probed. For instance, the $e^+e^- \to \gamma a~(a\to \gamma\gamma)$ and $Z \to a\gamma$ processes at LEP~\cite{Jaeckel:2015jla} have been exploited to search for ALPs. The $\gamma\gamma \to a \to \gamma\gamma$ 
searches have been utilized in electromagnetic PbPb collisions at the LHC by CMS and ATLAS~\cite{denterria2021collider}. The rare decay channels of Higgs boson $h \to Za ~(a\to \gamma\gamma)$ and $h \to a(\to \gamma\gamma)a(\to \gamma\gamma)$ at the LHC~\cite{Bauer:2017nlg} have also been utilized to probe the ALP-photon coupling $g_{a\gamma\gamma}$ versus the ALP mass $m_{a}$.
\item[(ii)] 
For an ALP with mass below the MeV scale, the cosmological and astrophysical observations~\cite{Raffelt:1990yz,Marsh:2015xka}, such as the Big Bang Nucleosynthesis (BBN), the Cosmic Microwave Background (CMB) and the Supernova 1987A, have already produced many constraints on ALP couplings. 
In addition, such light ALPs can become cold dark matter (DM)~\cite{Preskill:1982cy,Abbott:1982af,Dine:1982ah} and be detected by various astrophysical and terrestrial anomalies~\cite{Arias:2012az}, including the unexpected $X$-ray emission line around 3.5~keV~\cite{Jaeckel:2014qea} and the excess of electronic recoil events in XENON1T~\cite{Gao:2020wer}.
\item[(iii)] 
For an ALP in the mass range of MeV to GeV, it may make sizable contributions to low energy observables in particle physics and many searches in intensity frontiers~\cite{Izaguirre:2016dfi,Dolan:2017osp,Bauer:2019gfk,Banerjee:2020fue} which have been performed recently, such as the lepton flavor violating decays~\cite{Calibbi:2020jvd}, the rare meson decays~\cite{Izaguirre:2016dfi,Dolan:2017osp,Bjorkeroth:2018dzu,chakraborty2021heavy} and the ALP production in beam dumps experiments~\cite{Dobrich:2015jyk}. Besides, such an ALP has been proposed to explain the muon anomalous magnetic moment~\cite{Marciano:2016yhf,Bauer:2019gfk} and may also offer a plausible explanation for the Koto anomaly~\cite{Gori:2020xvq}. 
\item[(iv)] 
For the ALP mass in between 0.1 GeV and 10 GeV, recently, the Belle II~\cite{Abudin_n_2020} search for the process $e^{+}e^{-}\to\gamma a, a \to \gamma\gamma$ in the mass range 0.2 GeV $<m_a<$ 9.7 GeV using data corresponding to an integrated luminosity of (445$\pm$3)pb$^{-1}$.
\end{itemize}
Note that, a rather light ALP can be highly boosted, and thus the two photons from the ALP decay can be highly collimated and lead to an interesting ``diphoton-jet'' signature at the LHC. Distinguishing such diphoton-jets from the overwhelming QCD-jets and single photons is crucial for the ALP search at the LHC. In the literature there are some studies on these diphoton-jets by exploiting the jet substructure ~\cite{Ellis:2012zp,Allanach:2017qbs,Chakraborty:2017mbz,Sheff:2020jyw,Wang:2021uyb}. We will concentrate on the ALP that only couples to the electroweak gauge bosons with a mass range from a few hundred MeV to 10~GeV. Such a light ALP with couplings to gauge bosons has attracted much attention recently. Different from the existing studies~\cite{Gavela:2019cmq, Bauer:2017nlg}, we will  neglect the ALP-gluon coupling so that its dominant production channel at the LHC is the ALP-strahlung processes $pp \to Va$ ($V = W, Z$), as shown in Fig.~\ref{Feynman_diagrams}.

Besides the conventional analysis methods, the machine learning techniques have been used to search for new physics~\cite{Larkoski:2017jix,Ren:2017ymm,Abdughani:2018wrw,Ren:2019xhp,Collins:2019jip,Day:2019ucy,Abdughani:2019wuv,Abdughani:2020xfo,Feickert:2021ajf}. 
In our work we will exploit low-level jet information directly by utilizing machine learning (ML) and computer vision (CV) techniques, rather than using physics-inspired features, in order to not only improve the discrimination power, but also gain new insight into the underlying physical processes.
A convolutional neural network (CNN) will be designed to analyze these ``diphoton-jet'' events from the decay of ALPs based on the notion of jet-image which was first introduced in Ref.~\cite{Cogan:2014oua} and then studied in Refs.~\cite{Almeida:2015jua,de_Oliveira_2016,ATL-PHYS-PUB-2017-017,Lin:2018cin,Komiske_2018,Barnard_2017,Komiske_2017,Kasieczka_2017,Macaluso_2018,Li_2021,li2020attention,Lee_2019,collado2021learning,Du_2021}. In this way, the calorimeter is regarded as a camera and the jets are represented as images in which the pixel intensities are the energy depositions of the particles within the jet.
It was found in the literature that the CNN can provide the ability to learn rich high-level abstract features of jet images and greatly enhance the discrimination power. Based on the diphoton-jet tagging, we will further use another simple neural network (NN) to obtain optimal detection significance for the two ALP-strahlung processes in our study. 

This paper is organized as follows. In Section II, we describe the effective Lagrangian of ALP interactions and the simulation details. In Section III, we first introduce the jet-image pre-processing techniques, and then design a CNN architecture to identify the diphoton-jet events, and also build an NN to optimize the detection significance. In Section IV, we present and discuss the numerical results. Finally, we draw our conclusions in Section V.

\section{Model and simulation details}

\subsection{Model}
We consider an effective Lagrangian which consists of ALP interactions with electroweak gauge bosons up to dimension-5, given by~\cite{Brivio:2017ije}
\begin{eqnarray}
	\mathcal{L}_{\text{eff}} 
	&\supseteq &
 \frac{1}{2} (\partial^\mu a)(\partial_\mu a) - \frac{1}{2} m_a^2 a^2-  C_{BB} \frac{a}{f_a} B_{\mu\nu} \tilde{B}^{\mu\nu}
	-  C_{WW} \frac{a}{f_a} W^{i}_{\mu\nu} \tilde{W}^{\mu\nu,i},
\label{general-NLOLag-lin}
\end{eqnarray}
where $a$ represents the ALP field, $f_{a}$ is the ALP decay constant which is set to 1 TeV$^{-1}$ in this paper, and $V_{\mu\nu}$ and $\tilde{V}_{\mu\nu}$ represent the field strength for a SM gauge boson defined as $V_{\mu\nu} \equiv \partial_\mu V_\nu - \partial_\nu V_\mu$ and $\tilde{V}_{\mu\nu} \equiv \epsilon_{\mu\nu\rho\sigma}V^{\rho\sigma}$, with $W_{\mu\nu}$ and $B_{\mu\nu}$ being for $SU(2)_L$ and $U(1)_Y$, respectively. 

The third and fourth terms in Eq.~\ref{general-NLOLag-lin} induce the following dimensional couplings: $g_{a\gamma\gamma}$,  $g_{aWW}$, $g_{aZZ}$, $g_{a\gamma Z}$, which control the strength of the ALP's interaction with gauge bosons
\begin{eqnarray}
	g_{a\gamma\gamma} &=& \frac{4}{f_a} (C_{BB} \cos\theta_W^2 + C_{WW} \sin\theta_W^2),\\
	g_{aWW} &=& \frac{4}{f_a} C_{WW},\\
	g_{aZZ} &=& \frac{4}{f_a} (C_{BB} \sin\theta_W^2 + C_{WW} \cos\theta_W^2),\\
	g_{a\gamma Z} &=& \frac{8}{f_a}\sin\theta_W\cos\theta_W (C_{WW} - C_{BB}),
\end{eqnarray}
where $\theta_W$ is the Weinberg angle. For simplicity, we set $C_{WW} = C_{BB}$ in our study. In this work, we consider two ALP-strahlung production processes $p p \to W^\pm(\to \ell^\pm \nu)a(\to \gamma\gamma)$ and $p p \to Z(\to \ell^+\ell^-)a(\to \gamma\gamma)$ as our signals, in which $\ell^\pm$ denotes $e^\pm$ or $\mu^\pm$. The SM backgrounds of $Za$ signal are the $Z \gamma$ and $Z j$ processes, while the SM backgrounds of $W^\pm a$ signal are mainly the $W^\pm \gamma$ and $W j$ processes ($j$ stands for a light jet). Besides, the QCD di-jet events could also be an important SM background whenever one jet fragments into electromagnetic energy mostly and the other jet fragments into a ﬁnal state with an energetic isolated lepton. Although fake rate of jets is quite small, the large cross section of QCD di-jets makes its contribution non-negligible. In the LHC experiment, when $m_a \gtrsim 10$ GeV, ALP will decay into two well separated photons that are identified as $2\gamma$ events by the detector. On the other hand, if the ALP mass is lighter than a few hundred MeV, it will decay into highly collimated photon pairs that deposit energy in the electromagnetic calorimeter as a single photon. Between these two mass bounds, the two photons will be seen like a ``diphoton-jet''. In this case, we can use calorimeter towers as the pixels of the image and measure a jet as an image. Then we use machine learning techniques based on convolutional neural networks to discriminate diphoton-jets from single photons and QCD-jets.

\subsection{Simulation details}
For Monte Carlo signal simulations, we implement the effective Lagrangian of Eq.~\eqref{general-NLOLag-lin} in \textsf{FeynRules}~\cite{Alloul:2013bka} to generate the corresponding UFO model file. The parton level signal and background events are generated with \textsf{MadGraph5 aMC@NLO}~\cite{Alwall:2014hca} at the leading order. The following cuts for parton level event generation are employed: $p_{Tj}$ > 20 GeV, $|\eta_{j}|$ < 5.
We perform parton shower and fast detector simulations with
\textsf{Pythia8243}~\cite{Sjostrand:2014zea} and \textsf{Delphes 3.4.3}~\cite{deFavereau:2013fsa}. \textsf{FastJet 3.3.2}~\cite{Cacciari:2011ma} is used for jet clustering.
The NNPDF2.3QED parton distribution function (PDF) set~\cite{Ball:2013hta} is chosen in our calculations. Scale uncertainty is determined through independent restricted variation of the factorization scale $\mu_F$ and the normalization scale $\mu_R$. We use the built-in systematics tool in \textsf{MadGraph5 aMC@NLO} to evaluate the PDF and scale uncertainties. The cross sections of backgrounds are given in Table~\ref{uncertainty}. In $\textsf{Delphes}$, the stable hadrons such as neutrons and charged pions are assumed to deposit all their energy in the hadron calorimeter HCAL. For short-lived particles such as neutral pions decaying into a pair of photons are assumed to deposit all their energy in the electromagnetic calorimeter ECAL. For long-lived particles, such as Kaons and $\Lambda$ with $c\tau$ smaller than 10 mm, they are assumed to share their energy deposit between ECAL and HCAL with default fractions $f_\mathrm{ECAL}$=0.3 and $f_\mathrm{HCAL}$=0.7 according to the dominant decay products of those particles. The electron and muon are identified according to the default $\textsf{Delphes}$ CMS card, which are included in event selection. Only the photons and hadrons (including the hadronic decay products of tau) are used for jet reconstruction.

\begin{table}[ht]
\begin{center}\begin{tabular}{|c|c|c|c|c|c|c|c|c|c|c|}
\hline  ~~Backgrounds~~ & \tabincell{c}~~$\sigma_{tot}$[pb]~~ &  \tabincell{c}{scales[pb]} & \tabincell{c}{pdf[pb]}  \\
\hline \tabincell{c}{$t\bar{t}$} & 5.98$\times$10$^{2}$ & $ ^{+1.77\times10^{2} (29.6\%)}_{-1.28\times10^{2} (21.4\%)} $ & $ ^{+2.45\times10^{1} (4.1\%)}_{-2.45\times10^{1} (4.1\%)} $ \\
\hline \tabincell{c}{$tj$} & 2.04$\times$10$^{2}$ & $ ^{+1.87\times10^{1} (9.15\%)}_{-2.33\times10^{1} (11.4\%)} $ & $ ^{+2.82\times10^{0} (1.38\%)}_{-2.82\times10^{0} (1.38\%)} $ \\
\hline \tabincell{c}{$W^\pm\gamma$} & 9.91$\times$10$^{1}$ & $ ^{+1.04\times10^{1} (10.5\%)}_{-1.14\times10^{1} (11.5\%)} $ & $ ^{+3.09\times10^{0} (3.12\%)}_{-3.09\times10^{0} (3.12\%)} $ \\
\hline \tabincell{c}{$W^\pm j$} & 4.07$\times$10$^{4}$ & $ ^{+7.82\times10^{3} (19.2\%)}_{-6.80\times10^{3} (16.7\%)} $ & $ ^{+9.16\times10^{2} (2.25\%)}_{-9.16\times10^{2} (2.25\%)} $ \\
\hline \tabincell{c}{$jj$} & 2.49$\times$10$^{7}$ & $ ^{+7.94\times10^{6} (31.9\%)}_{-5.85\times10^{6} (23.5\%)} $ & $ ^{+4.10\times10^{5} (1.65\%)}_{-4.10\times10^{5} (1.65\%)} $\\
\hline \tabincell{c}{$Z\gamma$} & 7.55$\times$10$^{1}$ & $ ^{+7.85\times10^{0} (10.4\%)}_{-8.53\times10^{0} (11.3\%)} $ & $ ^{+2.31\times10^{0} (3.06\%)}_{-2.31\times10^{0} (3.06\%)} $\\
\hline \tabincell{c}{$Zj$} & 1.30$\times$10$^{4}$ & $ ^{+2.38\times10^{3} (18.3\%)}_{-2.07\times10^{3} (15.9\%)} $ & $ ^{+2.87\times10^{2} (2.2\%)}_{-2.87\times10^{2} (2.2\%)} $\\
\hline \end{tabular} \caption{The fiducial cross sections of the SM backgrounds with the theoretical uncertainties at 14 TeV LHC.}
\label{uncertainty}
\end{center}
\end{table}

 Based on the energy-flow algorithm~\cite{CMS-PAS-PFT-09-001}, the EflowPhotons, EflowNeutralHadrons and ChargedHadrons, which are composed of deposits in ECAL and HCAL, are clustered into jets using the anti-$k_t$ algorithm~\cite{Cacciari:2008gp} with $R_j = 0.4$. Only the leading jet in each event is retained for further analysis and it is required with $p_{T}$ > 50 GeV and $|\eta| < 2.5$. Besides, we also include the pile-up events to perform a pile-up robust analysis. The numerous soft QCD pile-up events are generated with \textsf{Pythia8} and then simulated by \textsf{Delphes}. In the CMS card we consider the average amount of pile-up events per bunch-crossing as 40. We take the default parametrization implemented in the CMS card to distribute the hard scattering events and pile-up events in time and $z$ positions randomly. The ground truth labels of the leading jets in signal events are ``diphoton-jet''. The leading jets in the $W^\pm a$ and $Za$ background events are labelled as ``single photon''. While the leading jets in other background events are labelled as ``QCD-jet''.

In terms of the effective ALP-photon coupling $g_{a\gamma\gamma}$, the decay width of the ALP is given by
\begin{equation}
    \Gamma = \frac{g^{2}_{a\gamma\gamma}m^{3}_{a}}{64\pi}
\end{equation}
Then the decay length in the limit of $E_{a}\gg m_{a}$ is approximated by~\cite{Dobrich:2015jyk}
\begin{equation}
   l_{a}=\beta\gamma\tau \approx \frac{64\pi E_{a}}{g^{2}_{a\gamma\gamma}m^{4}_{a}}\approx 40 \text{m} \times \frac{E_{a}}{\text{10 GeV}}(\frac{g_{a\gamma\gamma}}{10^{-5}\text{GeV}^{-1}})^{-2}(\frac{m_a}{100\text{MeV}})^{-4}
\end{equation}
Since the typical value of the transverse momentum of our ALP is in the range of 50 GeV to 1 TeV, we can have the decay length 0.1 mm $<l_a<$ 4.9 mm for $m_a=0.3$ GeV and $g_{a\gamma\gamma}=1$ TeV$^{-1}$. On the other hand, if the ALP is very light, the produced ALP at the LHC can be long-lived and may either be undetected (missing energy) or decay away from the primary vertex, other methods have to be utilized to observe such light ALP signature~\cite{Alimena_2020}.

In order to discriminate a diphoton-jet from a single photon and QCD-jet, we reorganize the information of jet constituents provided by the ECAL and HCAL as digital images which are the so-called jet images. In this work, the jet images are of $40\times40$ pixels resolution. They cover the area of [-0.4,0.4]$\times$[-0.4,0.4] in the $\eta\times\phi$ plane centering around the reconstructed jet axes. Thus, each pixel corresponds to $\Delta\eta \times \Delta\phi = 0.02 \times 0.02$, which matches the simulated CMS electromagnetic calorimeter granularity. As illustrated in Fig.~\ref{jet-image}, for each pixel, we sum up the transverse momentum separately for all the particles, for the charged hadrons, for the photons and for the neutral hadrons falling into the pixel as the pixel intensity in the four separate image channels. Therefore, each image channel is one jet observable.

\section{Jet tagging method}

\subsection{Jet image pre-processing}

To make the CNN learn highly discriminative features between signal and background events, we pre-process all the jet images by translation, rotation and normalization. Note that since the transverse momentum $p_{T}$ is invariant under longitudinal boosts, the pixel intensities in the four channels, which are given by the sum of the transverse momentum of all the particles, the charged hadrons, the photons and the neutral hadrons respectively inside the jet, are invariant under translation and rotation in $\eta$ and $\phi$.

\begin{itemize}

\item \textbf{Translation:} First of all, we define a new coordinate system in the $\eta-\phi$ plane centered at the jet. In this way, the coordinates of each particle $k$ inside the jet in the new coordinate system are given by 
\begin{eqnarray}
\eta_k^\prime = \eta_{k} - \eta_0,~~\phi_k^\prime = \phi_k - \phi_0.
\end{eqnarray}
where $(\eta_0,\phi_0)$ is the jet coordinates in the frame with the interaction point of p-p collision as the origin.

\item \textbf{Rotation:} The second step of pre-processing is rotating the jet image around the image center. We define the ``barycenter'' of a jet image as
\begin{eqnarray}
    \eta_C = \frac{1}{p_{T}}\sum_k \eta'_k p_{T_k},
    \quad
    \phi_C = \frac{1}{p_{T}}\sum_k \phi'_k p_{T_k},
\end{eqnarray}
which is the weighted sum of the transverse momenta of all the constituents inside the jet, where $p_{T} = \sum_k p_{T_k}$ is the total transverse momentum of the jet and $p_{T_k}$ is the transverse momentum of the $k$-th particle inside the jet. Then, we rotate the whole jet image such that the ``barycenter'' of the jet is in the 12 o'clock position:
\begin{eqnarray}
    \tan\theta = \phi_C/\eta_C
\end{eqnarray}
\begin{eqnarray}
    \eta_k^{\prime\prime} = \eta_k^\prime\cos\theta-\phi_k^\prime\sin\theta,
    \phi_k^{\prime\prime} = \eta_k^\prime\sin\theta+\phi_k^\prime\cos\theta.
\end{eqnarray}

\item \textbf{Normalization:} The last step is normalization.
In order to make the neural network learn more efficiently, we scale the pixel intensity of each channel to [0,1] via dividing by a constant. In this way, the absolute energy scale dependence of jet images can be removed. It allows for comparisons of jet images with different energies. We take different normalization constants for different channels. Most of the pixel intensities of diphoton-jet images and single photon images are carried by the the pair of collimated photons and the single photon with transverse momentum of tens of GeV. Therefore for the channel given by the sum of the transverse momentum of all the particles and the channel given by the sum of the transverse momentum of all the photons, the normalization constant is set to 100. The charged hadrons and neutrons distributions in all the jet images samples are diffuse and the transverse momentum of most of them are less than 10 GeV. Therefore for the channel given by the sum of the transverse momentum of all the charged hadrons or all the neutral hadrons, the normalization constant is set to 10.

\end{itemize}

\subsection{Jet-tagging neural network}

\begin{figure}[ht]
    \centering
    \includegraphics[width=16cm,height=10cm]{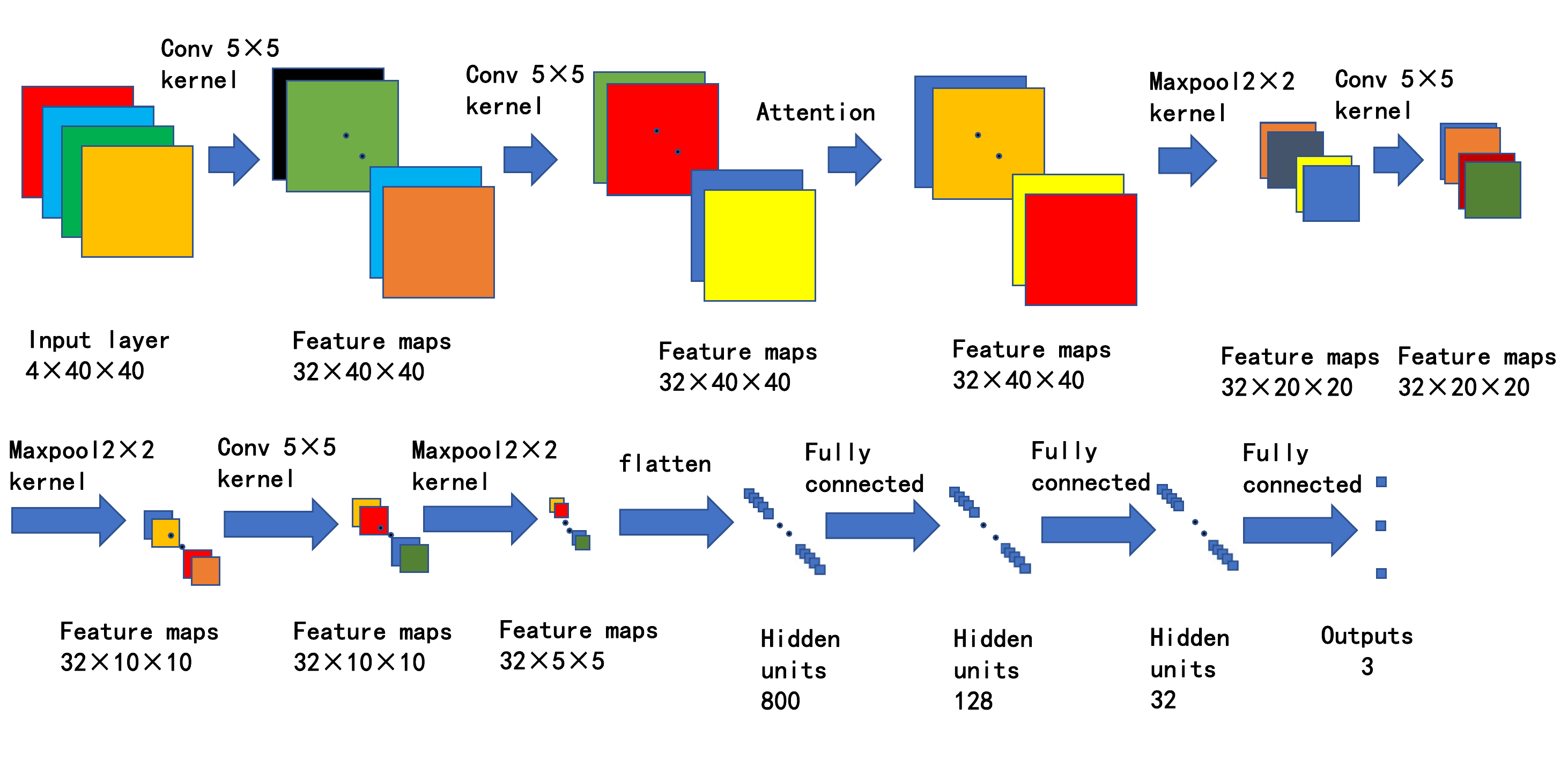}
    \caption{Illustration of the jet-tagging CNN}
    \label{cnn1}
\end{figure}

An illustration of the architecture of the jet-tagging CNN is shown in Fig.~\ref{cnn1}. The input is a jet-image of $40\times40$ pixels with four channels. Followed by the input layer, a normal convolutional layer and a depthwise convolutional layer are designed to extract pixel-level feature map of the jet-image. The kernel size of the normal convolutional layer is $5\times 5$ and the number of kernels is 32.
The inner structure of the depthwise convolutional layer~\cite{chollet2017xception} is given in Fig~\ref{depthwise}, where a depthwise convolution of $5 \times 5$ kernel is first applied on each channel of the input feature map, and then a $1 \times 1$ pointwise convolution is applied to aggregate the features along channel direction. Besides, the leaky ReLU~\cite{maas2013rectifier} is adopted as the activation after both the normal convolutional layer and the depthwise convolutional layer.

\begin{figure}[ht]
    \centering
    \subfigure[The overall structure of the attention block]{
        \includegraphics[width=15cm,height=6cm]{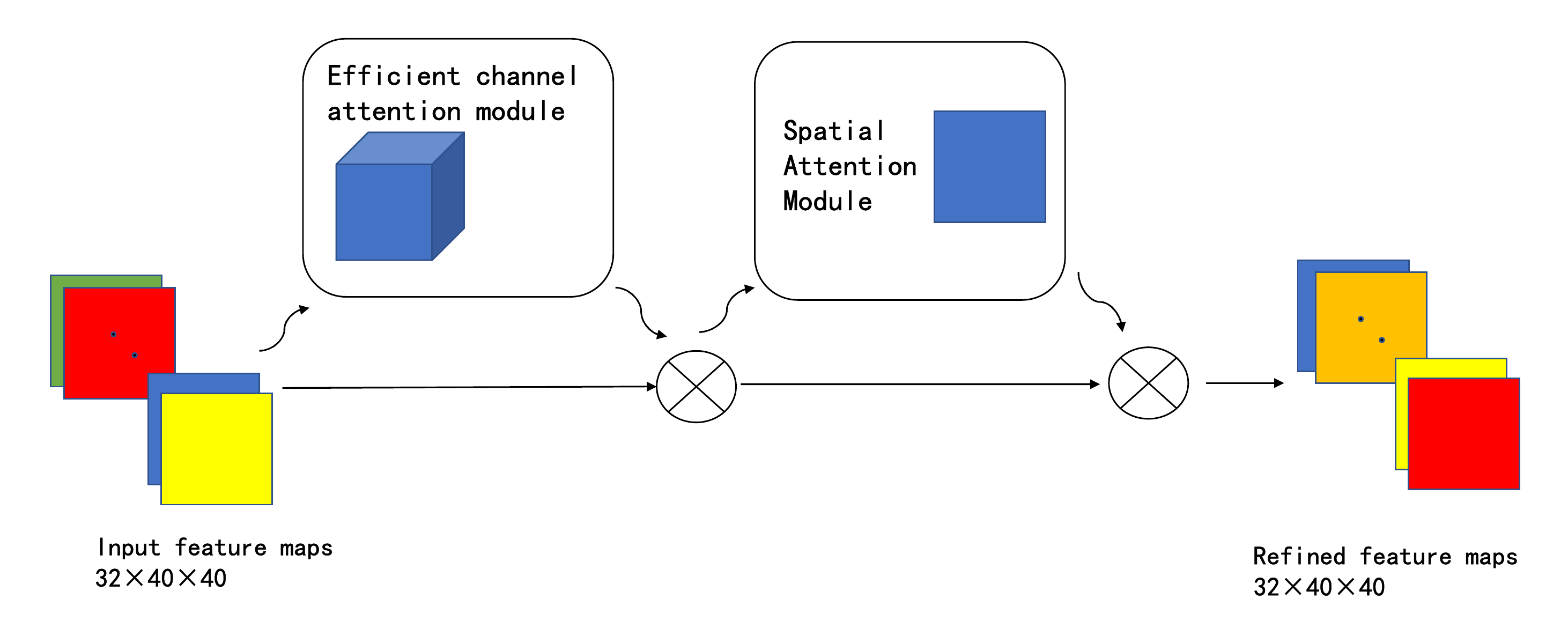}
        \label{attention-overall}
    }
    \subfigure[The efficient channel attention (ECA) module]{
        \includegraphics[width=15cm,height=6cm]{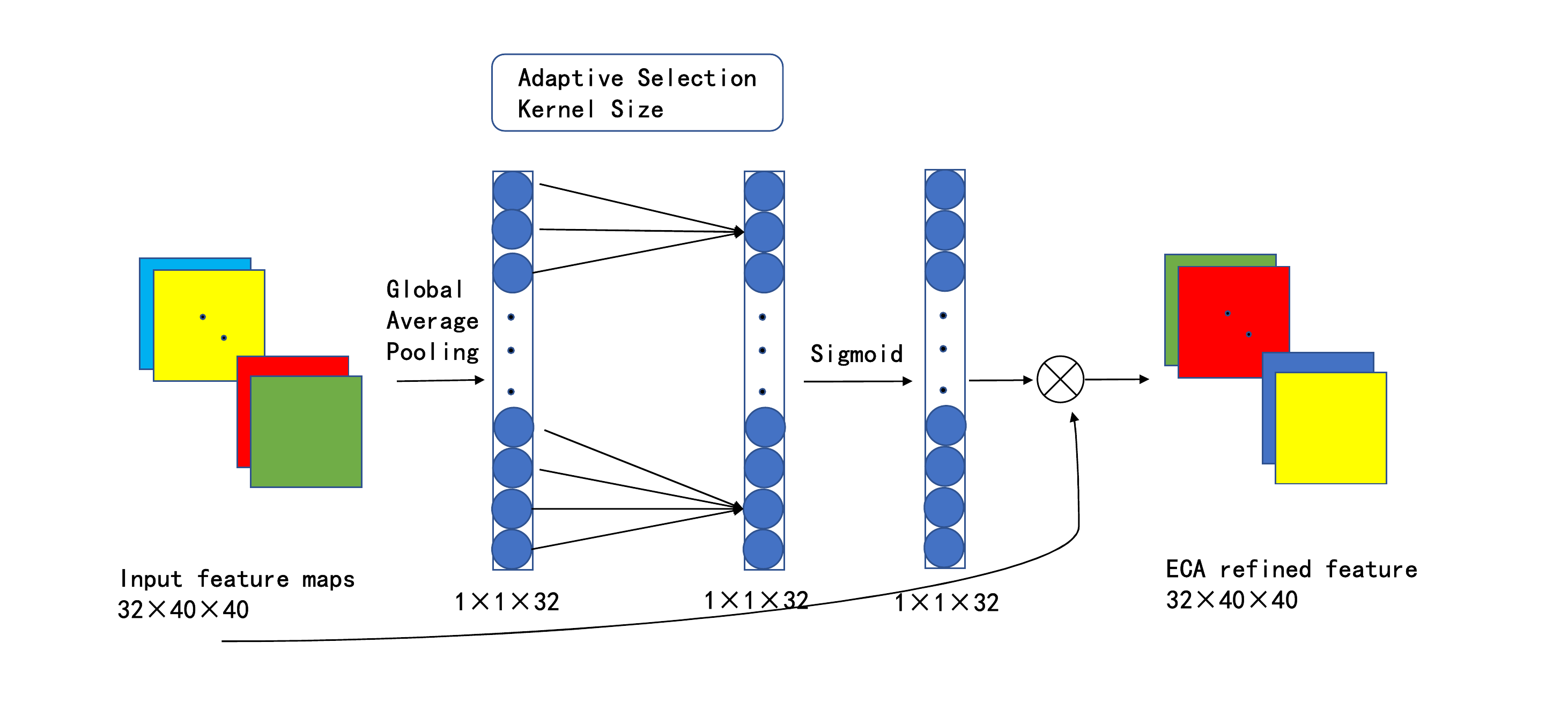}
        \label{attention-eca}
    }
    \subfigure[The spatial attention module]{
        \includegraphics[width=15cm,height=6cm]{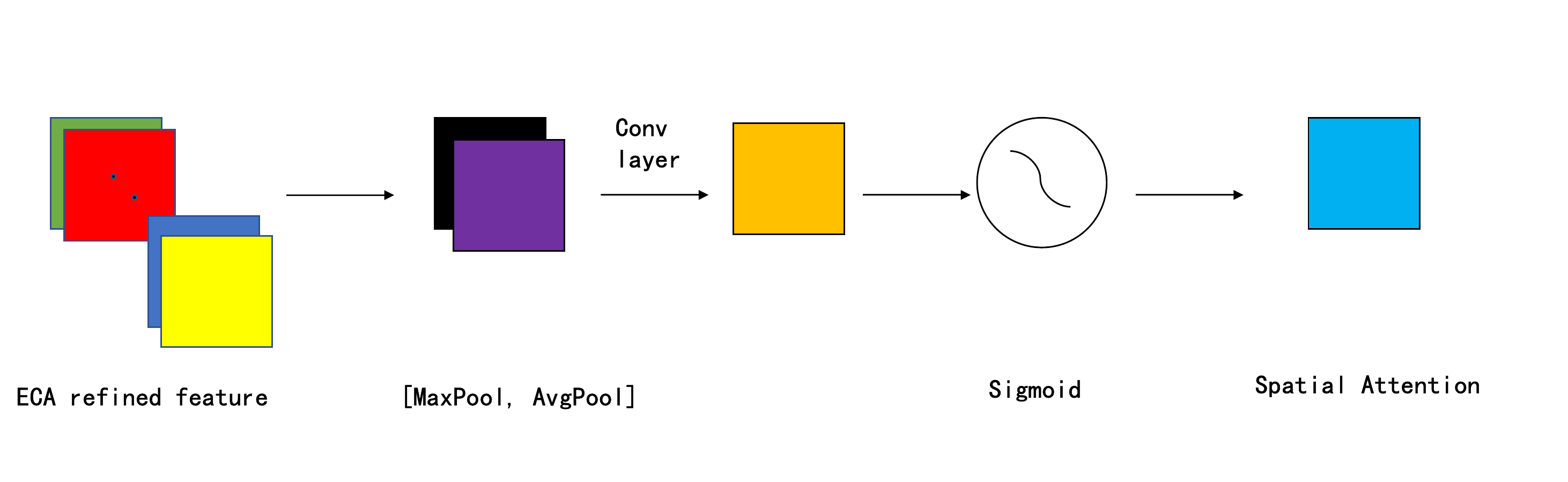}
        \label{attention-spatial}
    }
    \caption{The attention block designed in our model.}
    \label{attention}
\end{figure}

Then, we add an attention block to weight the feature map, as shown in Fig.~\ref{attention-overall}, so as to make the CNN concentrate on the significant features distinguishing signals from backgrounds. The resulting feature map obtained from the previous layer is taken as the input of an efficient channel attention (ECA) module \cite{wang2020ecanet}. As shown in Fig.~\ref{attention-eca}, by independently averaging all the pixels values of each feature map with each average value representing the corresponding feature channel, the global average pooling results in a feature vector whose dimension equals to the number of input channels which is 32 in our designed model. Then, a one-dimension convolution of kernel size 5 is adopted to aggregate the local information in the feature vector and the result is activated by a sigmoid function to generate the weights on each feature channel. Finally, the input feature map is weighted by multiplying each feature channel by its corresponding weight. As a result, the ECA module suppresses less important feature channels. In short, ECA attention is computed as 
\begin{align}\label{fun_eca1}
\omega_{m} = \sigma\bigg(\sum_{n=1}^{k}w_{mn}y_{mn}\bigg),\,\,y_{mn} \in \Omega_{mk},
\end{align}
where $\Omega_{mk}$ indicates the set of $k$ adjacent channels of $y_{m}$ and $\sigma$ denotes the sigmoid function. The weight of $m$-th channel is calculated by considering the local cross-channel interaction between $y_m$ and its $k$ neighbors. In this way, the ECA module first captures the local information which is exchanged quickly across channels through each channel and its $k$ neighbors. Then the global cross-channel interactive information is captured by fast one-dimensional convolution of size $k$. This size $k$ represents the coverage of the local cross-channel interaction, i.e., how many neighbors participate in attention prediction of one channel. In order to avoid manual tuning of $k$ via cross-validation, $k$ is determined adaptively.

Secondly, the weighted feature map obtained from the ECA module is further fed into a spatial attention module \cite{woo2018cbam}. As shown in Fig.~\ref{attention-spatial}, both a global max pooling and a global average pooling are applied on the input feature map along the channel direction to generate two aggregated feature channels. Then, the spatial weights are generated by applying a convolution of kernel size $3 \times 3$ and a sigmoid activation. Finally, the input feature map is multiplied by the spatial weights pixel-wisely. In short, the spatial attention is computed as 
\begin{equation}\label{eq:forth}
\begin{split}
    \mathbf{M_s}(\mathbf{F})&=\sigma(f^{3\times 3}([AvgPool(\mathbf{F}); MaxPool(\mathbf{F})]))\\
    &=\sigma(f^{3\times 3}([\mathbf{F^{s}_{avg}}; \mathbf{F^{s}_{max}}])),
\end{split}
\end{equation}
where \(\sigma\) denotes the sigmoid function and \(f^{3\times 3}\) represents a convolution operation with the filter size of $3\times 3$. \(\mathbf{F^{s}_{avg}}\in \mathbb{R}^{1\times 40\times 40}\) and \(\mathbf{F^{s}_{max}}\in \mathbb{R}^{1\times 40\times 40}\) represent the two 2D maps generated by the two pooling operations.

Next, we apply a stack of $2\times2$ maxpooling layers and $5\times5$ depthwise convolutional layers. All the feature maps have 32 channels and are activated by a leaky ReLU.

Afterwards, we flatten the final feature map into a single vector and feed it into a fully connected network with two hidden layers. The hidden layers have 128 and 32 neurons, respectively, with a leaky ReLU activations. In the output layer, we apply the SoftMax activation to generate three class probabilities for the diphoton-jet, the single photon and the QCD-jet, respectively. The sum of the three probabilities is one. 

To optimize the model parameters in the jet-tagging CNN, we choose the cross-entropy as the loss function. The CNN is trained using the Adam \cite{Kingma:2014vow} optimizer with a constant learning rate of 0.001 based on the gradients calculated on a mini-batch of 64 training examples. The network is trained up to 100 epochs, and we adopt the early-stopping technique to prevent over-fitting.

\subsection{Jet selection neural network}

\begin{figure}[ht]
    \centering
    \includegraphics[width=6cm,height=6cm]{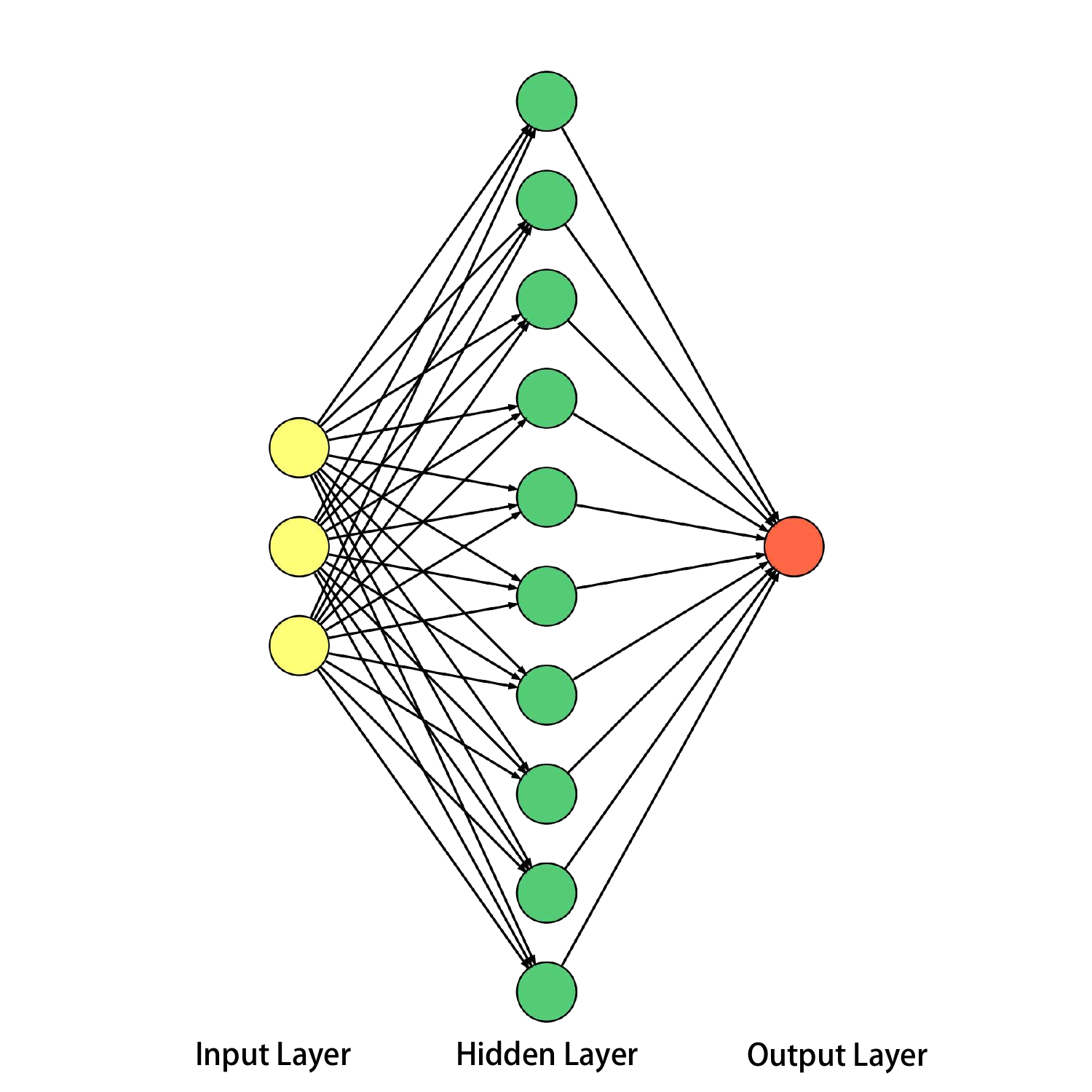}
    \caption{Architecture of the selection neural network.}
    \label{cnn2}
\end{figure}

\begin{figure}[ht]
\begin{center}
\includegraphics[width=15cm,height=8cm]{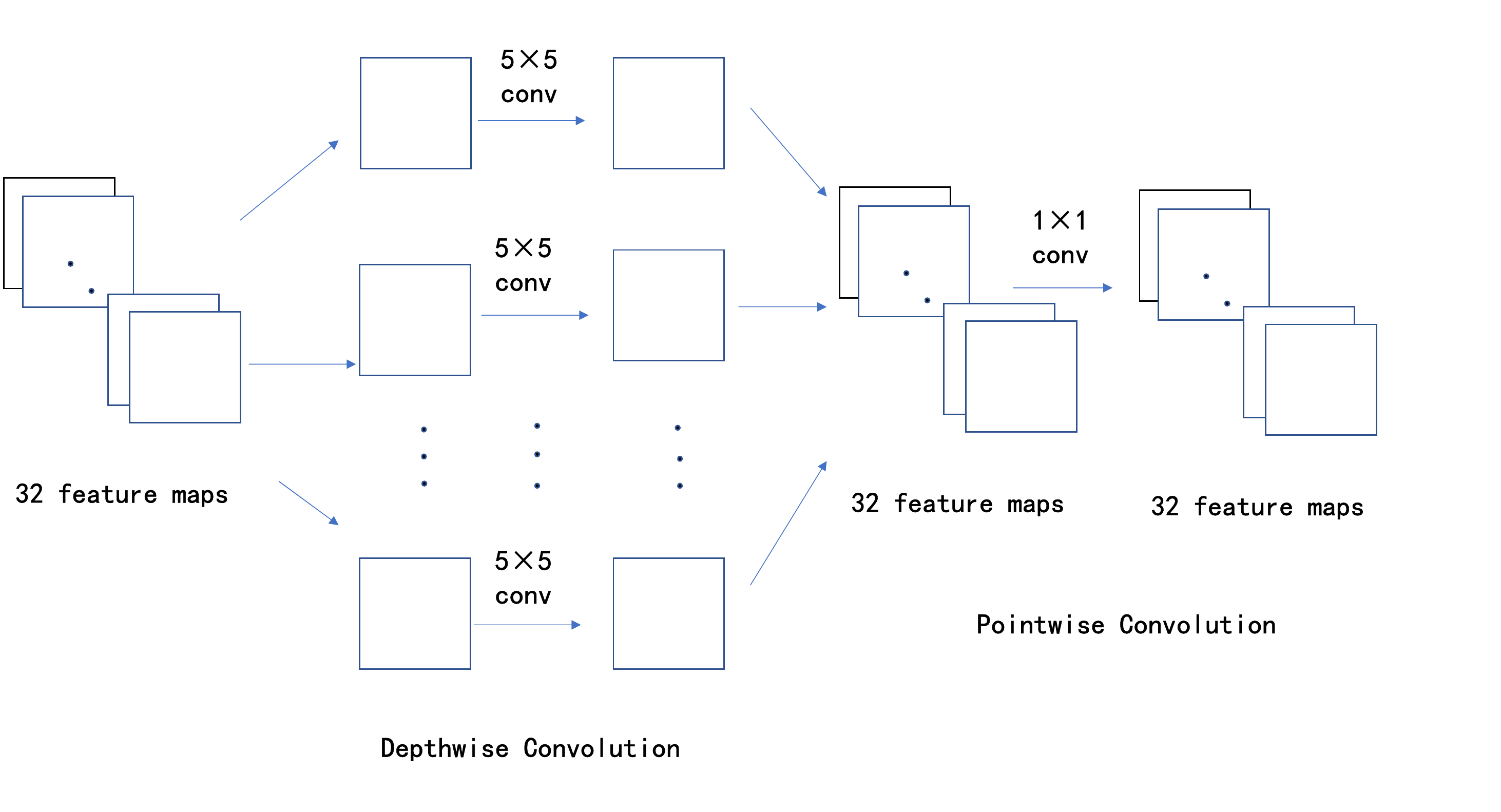}
\caption{Illustration of the depthwise convolution layer and the pointwise convolution layer.}
\label{depthwise}
\end{center} 
\end{figure}

\begin{figure}[!h]
    \centering
    \subfigure[diphoton-jet]{
        \includegraphics[height=10cm,width=4.2cm]{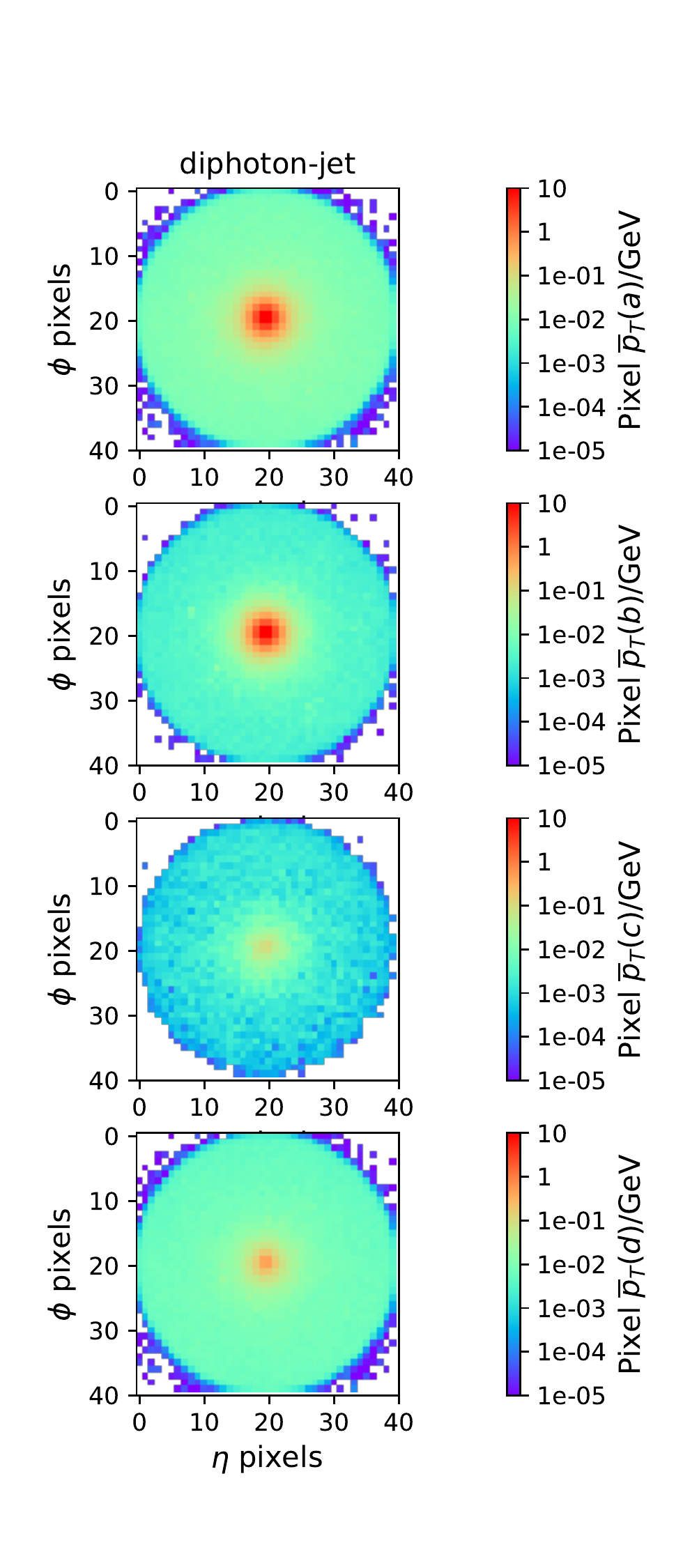}
    }
    \subfigure[single photon]{
        \includegraphics[height=10cm,width=4.2cm]{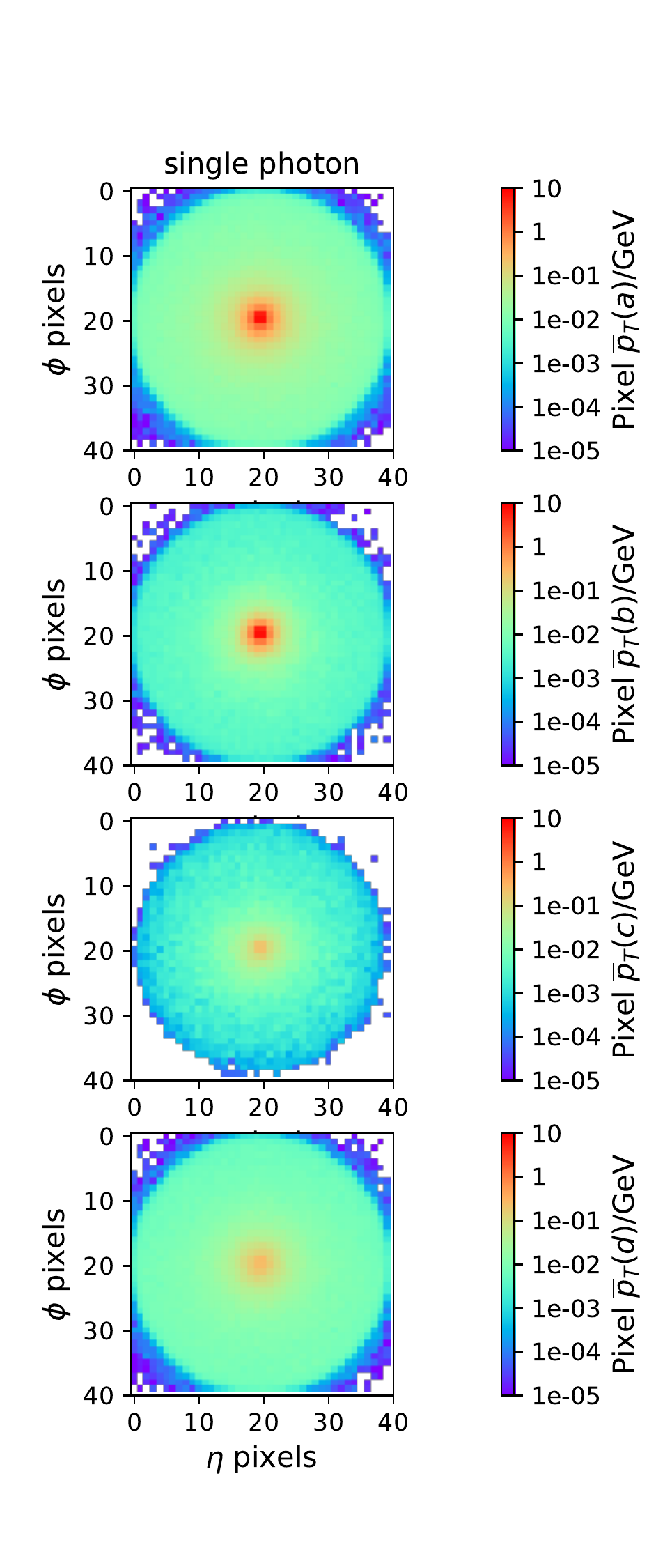}
    }
    \subfigure[QCD-jet]{
        \includegraphics[height=10cm,width=4.2cm]{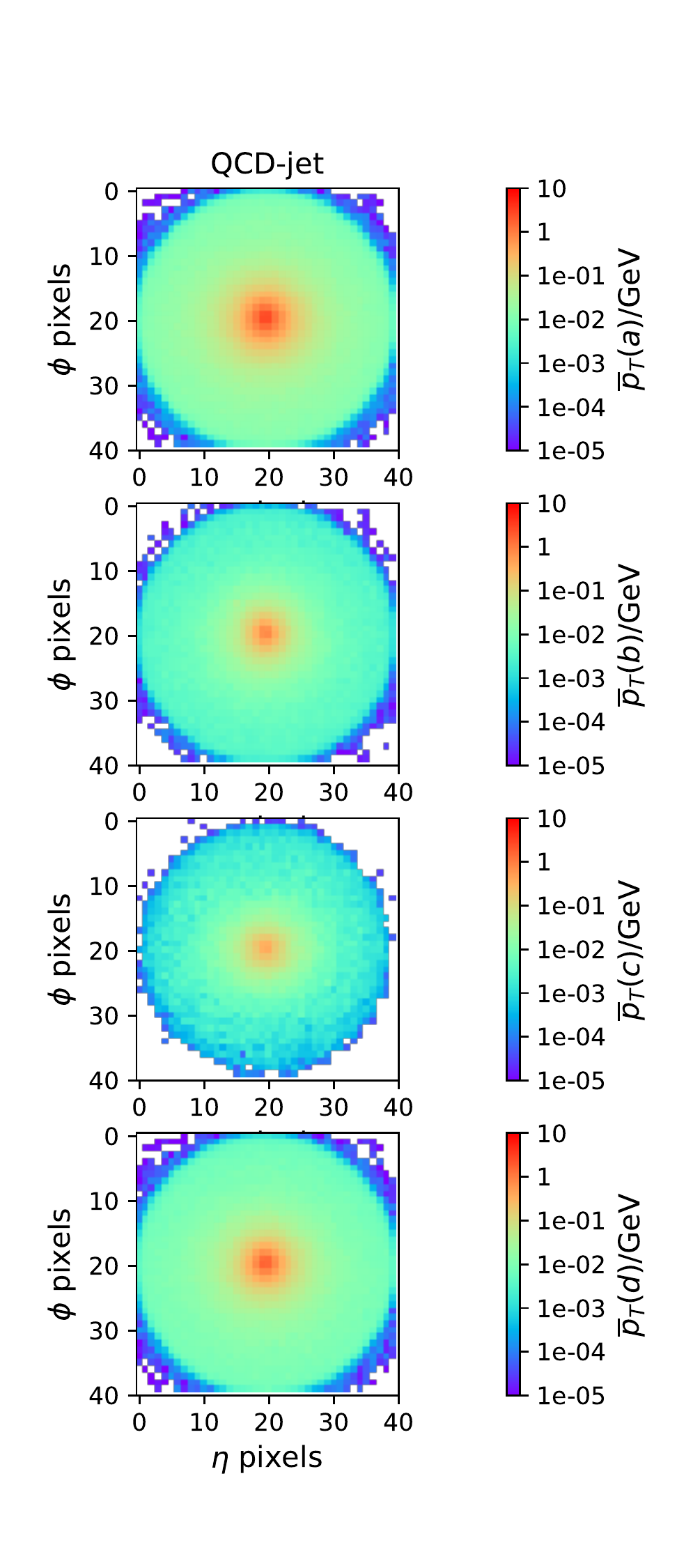}
    }
    \caption{The jet images for three kinds of jets after applying the translation and rotation steps of the pre-processing. The three kinds of jets are the leading jets of the benchmark signal events with $g_{a\gamma\gamma}=0.64$ TeV$^{-1}$ and $m_{a} = 3$~GeV, the leading jets of the W$^\pm \gamma$ background events and the leading jets of the W$^\pm j$ background events after basic selection.} The pixel intensities of the four channels, defined by the sum of the transverse momentum of all the particles, the photons, the charged hadrons and the neutral hadrons, are labeled as $p_{Ta}$, $p_{Tb}$, $p_{Tc}$ and $p_{Td}$, respectively.
    \label{jet-image}
\end{figure}

Detection significance depends on the event selection efficiency. In order to obtain an optimal detection significance, another simple neural network is built to learn an optimal selection cut on jet tagging probabilities. As shown in Fig.~\ref{cnn2}, it takes the three jet-tagging probabilities obtained from the jet-tagging neural network as input. A hidden layer with 10 neurons and a leaky ReLU activation is used to enhance the non-linearity of the optimized cut. The output layer is activated by a sigmoid function. This network predicts the probability of the jet selected as a signal.

Because the detection significance is calculated by counting the signal and background events, it cannot be used directly to build the loss function of the optimization. Since the loss function should be mathematically differentiable, we first approximate the signal and background event count using the jet selection probabilities obtained from the neural network for all events. Denoting the jet selection probability for event $i$ as $p_i$, the jet selection efficiency can be estimated by
\begin{equation}
    \varepsilon_\mathrm{pj} = \frac{\sum_\mathrm{pj} p_\mathrm{pj}}{\sum_\mathrm{pj} 1},
    \varepsilon_\mathrm{p} = \frac{\sum_\mathrm{p} p_\mathrm{p}}{\sum_\mathrm{p} 1},
    \varepsilon_\mathrm{j} = \frac{\sum_\mathrm{j} p_\mathrm{j}}{\sum_\mathrm{j} 1} ,
\end{equation}
where "$pj$", "$p$" and "$j$" represent the diphoton-jet signal, the single photon background and the QCD-jet background events, respectively. The sum runs over all the corresponding events. Then the signal and background event count $\mathcal{S}$ and $\mathcal{B}$ can be approximated by
\begin{align}
    \mathcal{S} &\approx \sigma_S \varepsilon_{pj} \mathcal{L} , \\
    \mathcal{B} &\approx (\sigma_p \varepsilon_p + \sigma_j \varepsilon_j) \mathcal{L},
\end{align}
where $\sigma_S, \sigma_p$ and $\sigma_j$ represent the cross sections of the diphoton-jet signal, the single photon background and the QCD-jet background after the basic selection which will be described in the next section. Then, the simple detection significance formula can be expressed as
\begin{equation}
    Z = \frac{\mathcal{S}}{\sqrt{\mathcal{B}}} = \frac{\sigma_S \varepsilon_{pj} \mathcal{L}}{\sqrt{(\sigma_p \varepsilon_p + \sigma_j \varepsilon_j) \mathcal{L}}}
\end{equation}
We take $-Z$ as the loss function to optimize the NN model weights. Then, after optimization, we take 0.5 as the selection threshold to count the number of diphoton-jets, single photons and QCD-jets (whose selection probabilities are greater than 0.5) and use the Poisson formula
\begin{equation}
    Z = \sqrt{2\left[(\mathcal{S}+\mathcal{B}){\rm ln}(1+\mathcal{S}/\mathcal{B})-\mathcal{S}\right]}
\end{equation}
to accurately evaluate the detection significance.
In our analysis, we use the Adam optimizer with a learning rate of 0.02 to optimize the model weights based on the gradients calculated over 5000 epoch.

Both the CNN and the simple NN were implemented in the deep learning framework of PyTorch~\cite{NIPS2019_9015} with GPU acceleration. Xavier-uniform initialization~\cite{glorot11a} is used to initialize the model weights and the model biases are initialized to zero.

\section{Analysis and results}

For the ALP-strahlung production process $p p \to W^\pm a$, the final states are identified as one lepton and one diphoton-jet. The main SM backgrounds are from the QCD di-jet, $W^\pm j$, $W^\pm \gamma$, $t\bar{t}$ and $tj$ productions. According to the above analysis, we adopt a basic selection criteria to select signal events in our analysis: (i) There is exactly one lepton (electron or muon) with $p_\text{T} > 20$ GeV and $|\eta| < 2.5$;
(ii) The hardest jet is required to have $p_\text{T} > 50$ GeV and $|\eta| < 2.5$ (since our signal contains a hard diphoton-jet).

For the ALP-strahlung process $p p \to Z a$, the final states are marked by an opposite-sign and 
same-flavor charged lepton pair and a diphoton-jet. 
The main SM backgrounds are dominated by the productions of $Z\gamma$ and $Zj$. We distinguish the signal and the background by imposing a basic selection criteria:
(i) There are exactly two oppositely charged leptons with $p_\text{T} > 20$ GeV and $|\eta| < 2.5$;
(ii) The invariant mass of the oppositely charged lepton pair with same flavor is required to be in the range of 70 GeV < $m_{ll}$ < 110 GeV;
(iii) The hardest jet is required to have $p_\text{T} > 50$ GeV and $|\eta| < 2.5$ (for the reason stated in the above). 

According to the L1 trigger menu of the CMS with tracking information~\cite{CMSrates}, the current offline threshold of the single electron (muon) $+$ jet are 23 GeV (16 GeV) and 66 GeV, respectively. The offline threshold of the single electron $+$ photon are 22 GeV and 16 GeV, respectively. These will be able to collect all events of interest for our study. The higher event rates and event sizes at the HL-LHC will be a challenge for the trigger and data acquisition systems. With a complex series of upgrades including the installation of new detectors and the replacement of ageing electronics, we expect the future trigger menu could be comparable with that of Run-2.

\begin{table}[ht]
\begin{center}
\resizebox{\textwidth}{25mm}{
\begin{tabular}{|c|c|c|c|c|c|c|c|c|c|c|}
\hline  basic selection & \tabincell{c}{signal} &  \tabincell{c}{$jj$} & \tabincell{c}{$W^\pm\gamma$} & \tabincell{c}{ $W^\pm j$}  &\tabincell{c}{ $t\bar{t}$} & \tabincell{c}{$tj$}   \\
\hline  \tabincell{c}{No Cut} & $4.6\times10^{-1}$ & $2.5\times10^7$ & $9.9\times10^1$ & $4.1\times10^4$ & $6.0\times10^2$ & $2.0\times10^2$ \\
\hline  \tabincell{c}{1 lepton with \\$p_\text{T} > 20$ GeV \\and $|\eta| < 2.5$} & $6.0\times10^{-2}$ & $1.9\times10^5$ & $1.2\times10^1$ & $4.4\times10^3$ & $1.5\times10^2$ & $2.9\times10^1$ \\
\hline  \tabincell{c}{The hardest \\jet with \\$p_\text{T} > 50$ GeV \\and $|\eta| < 2.5$} & $4.0\times10^{-2}$ & $1.3\times10^5$ & $2.5\times10^0$ & $1.6\times10^3$ & $1.4\times10^2$ & $1.9\times10^1$ \\
\hline \end{tabular}}
\caption{The basic selection cut-flow of the cross sections (in units of pb) for the ALP-strahlung production process $pp \to W^\pm a$ with $m_{a}$=3 GeV at the 14 TeV LHC where $g_{a\gamma\gamma}$ are set to 0.64 TeV$^{-1}$. As a comparison, the corresponding results of the backgrounds are also listed. }
\label{Cutflow1}
\end{center}
\end{table}
\begin{table}[ht]
\begin{center}\begin{tabular}{|c|c|c|c|c|c|c|c|c|c|c|}
\hline  ~~basic selection~~ & \tabincell{c}~~signal~~ &  \tabincell{c}{$Z\gamma$} & \tabincell{c}{$Zj$}  \\
\hline \tabincell{c}{No Cut} & $1.7\times10^{-1}$ & $7.6\times10^1$ & $1.3\times10^4$ \\
\hline \tabincell{c}{2 leptons with \\$p_\text{T} > 20$ GeV \\and $|\eta| < 2.5$} & $4.2\times10^{-3}$ & $1.9\times10^0$ & $2.8\times10^2$ \\
\hline \tabincell{c}{oppositely charged lepton\\ pair with same flavor and\\ 70 GeV < $m_{ll}$ < 110 GeV} & $4.1\times10^{-3}$ & $1.9\times10^0$ & $2.8\times10^2$ \\
\hline \tabincell{c}{The hardest jet \\with $p_\text{T} > 50$ GeV \\and $|\eta| < 2.5$} & $2.9\times10^{-3}$ & $4.4\times10^{-1}$ &~~ $1.0\times10^2$ ~~\\
\hline \end{tabular} \caption{Same as Table 1, but for the ALP-strahlung production process $pp \to Za$ and the corresponding backgrounds.}
\label{Cutflow2}
\end{center}
\end{table}

As an example, we consider a benchmark signal point with $g_{a\gamma\gamma}=0.64$ TeV$^{-1}$ and $m_{a} = 3$~GeV. The cut-flow of our basic selection for the ALP-strahlung production processes $pp \to W^\pm a$ and $pp \to Z a$ and the main backgrounds are given in Tables~\ref{Cutflow1} and \ref{Cutflow2}. Next, we use the signal and background events after basic selection to train the CNN.
During the 100 training epochs, the model with the minimal validation loss is chosen as the best model.
The size of the training set and validation sets are both 150k.

After applying the above translation and rotation in the preprocess, in Fig.~6, we present the jet images of the diphoton-jet from the ALP, the single photon and the QCD-jet. The benchmark signal events are generated for the ALP with $g_{a\gamma\gamma}=0.64$ TeV$^{-1}$ and $m_{a} = 3$~GeV. The single photon and QCD-jet are taken from $W^\pm+\gamma$ and $W^\pm+j$ background events, respectively. The pixel intensities $\bar{p}_{T}(a)$, $\bar{p}_{T}(b)$, $\bar{p}_{T}(c)$ and $\bar{p}_{T}(d)$ (namely four image channels) correspond to the averages of the transverse momentum of all the particles, the photons, the charged hadrons, and the neutral hadrons falling in each pixel over the total number of events, respectively. As shown in Fig.~6, the diphoton-jet and the single photon events have much higher $p_{Ta}$ and $p_{Tb}$ than the QCD-jet events. Meanwhile, the QCD-jet background has much higher $p_{Tc}$ and $p_{Td}$ than the diphoton-jet and the single photon events. In the four image channels, the spread of the pixel intensity of the QCD-jet events around each image center is wider than that of the diphoton-jet and the single photon events.

\begin{figure*}[ht]
    \centering
    \includegraphics[width=4.9cm]{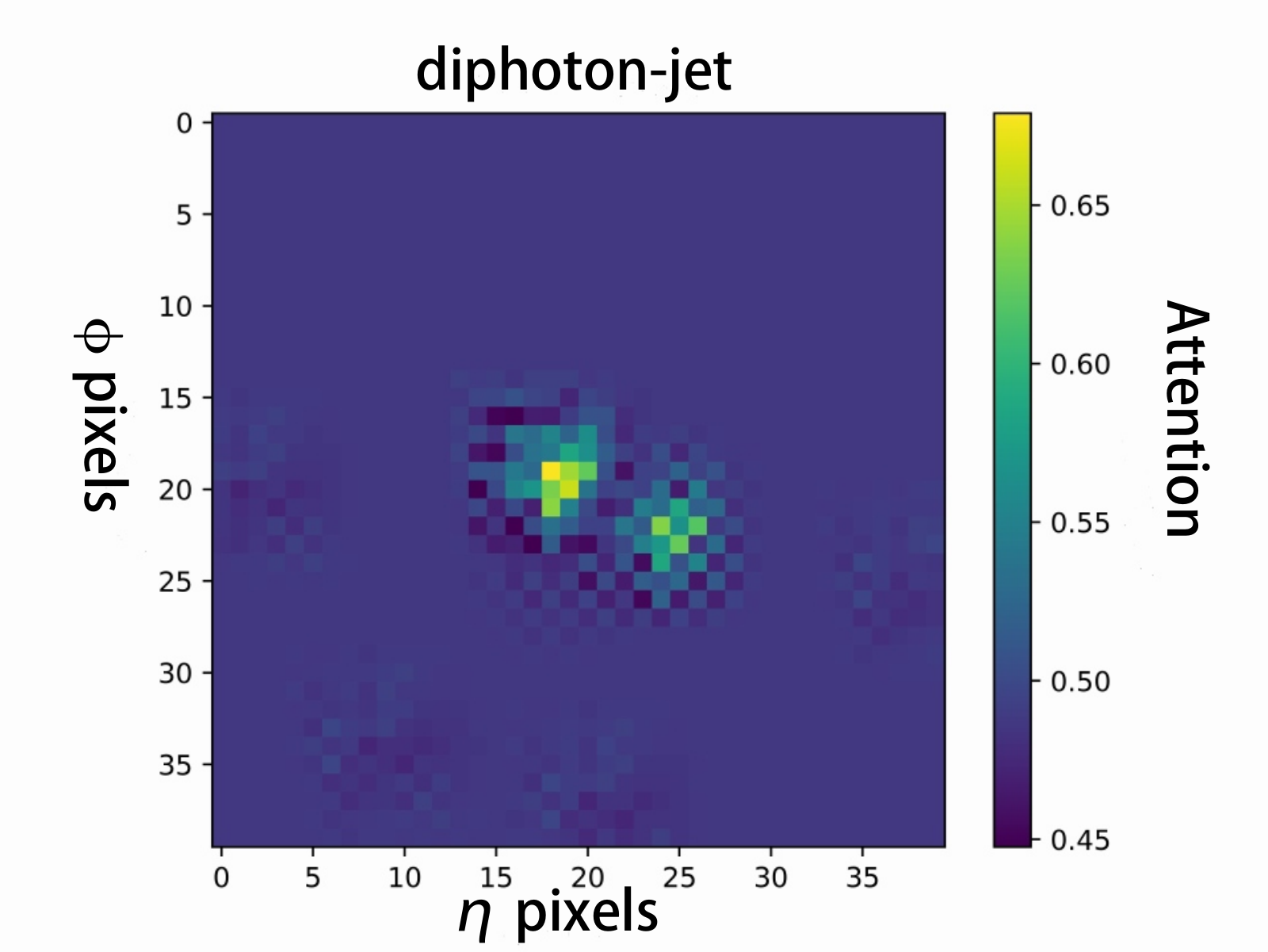}
    \includegraphics[width=4.9cm]{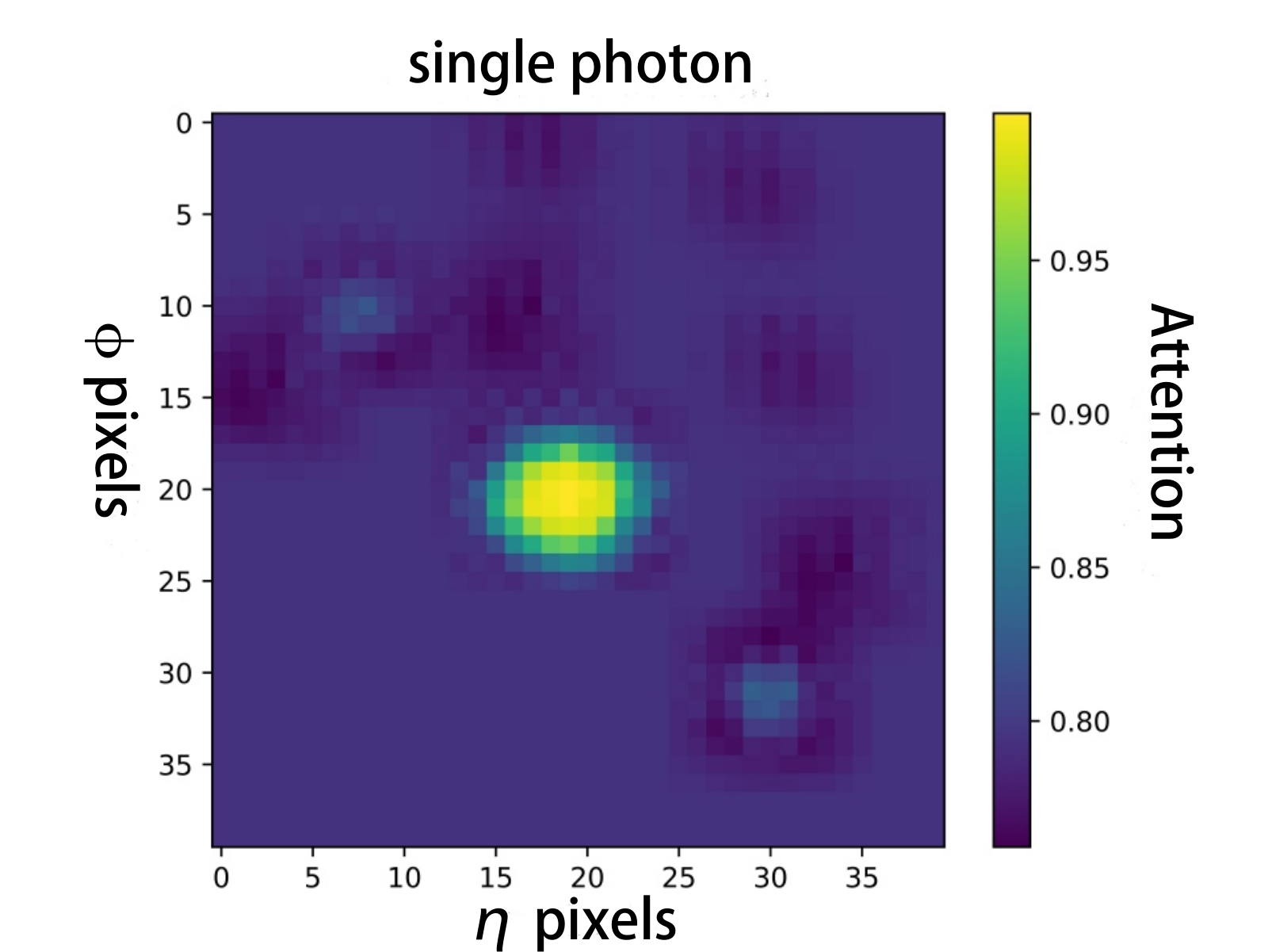}
    \includegraphics[width=4.9cm]{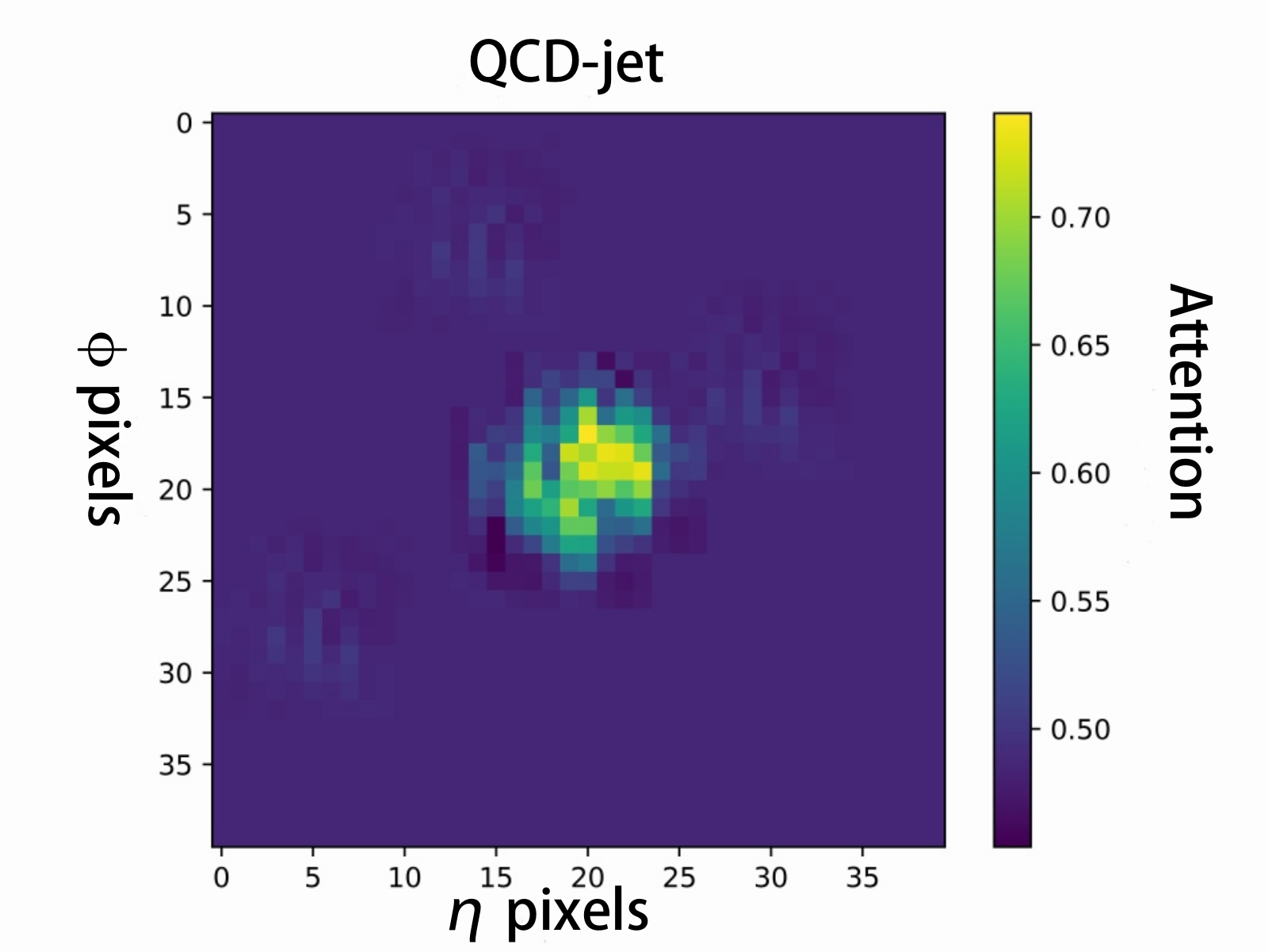}
    \caption{Attention obtained by the training samples for the QCD-jets, single photons and diphoton-jets when $m_{a}=3$ GeV.}
    \label{attention_images}
\end{figure*}

Fig.~\ref{attention_images} shows the attention images for a QCD-jet, a single photon and a diphoton-jet samples, respectively, when $m_{a} = 3$~GeV. It shows that the network can learn to automatically extract the most distinguishable image regions and pay more attention to them. For a diphoton-jet sample, the network focuses on two image regions that contain a pair of collimated photons, and the leading subjet gets relatively stronger attention. For a single photon sample, the network pays strong attention to the pixel where the leading subjet is located. And the attention is weak and more diffuse for a QCD-jet sample.

\begin{figure*}[ht]
\begin{center}
\includegraphics[width=4.5cm]{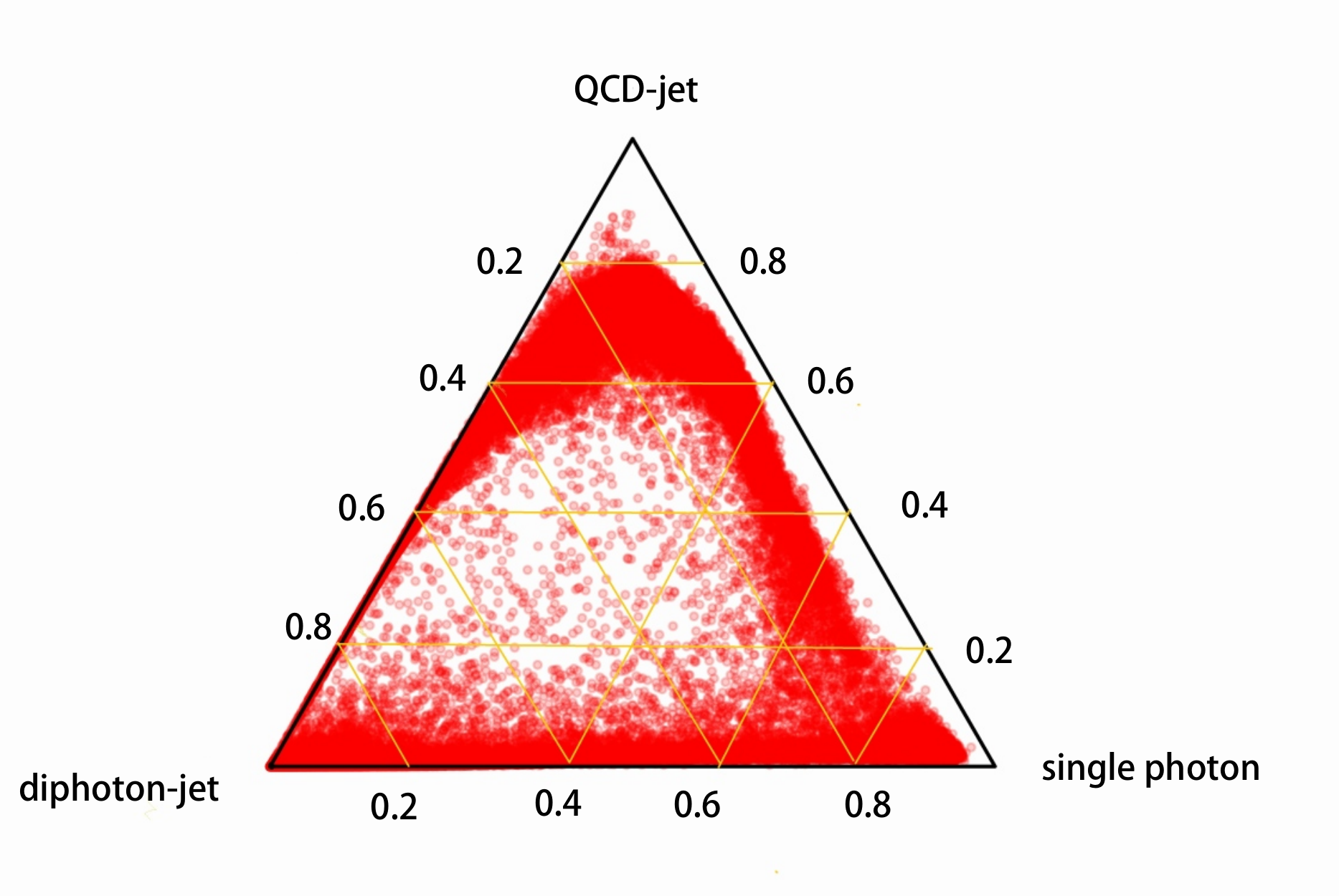}
\includegraphics[width=4.5cm]{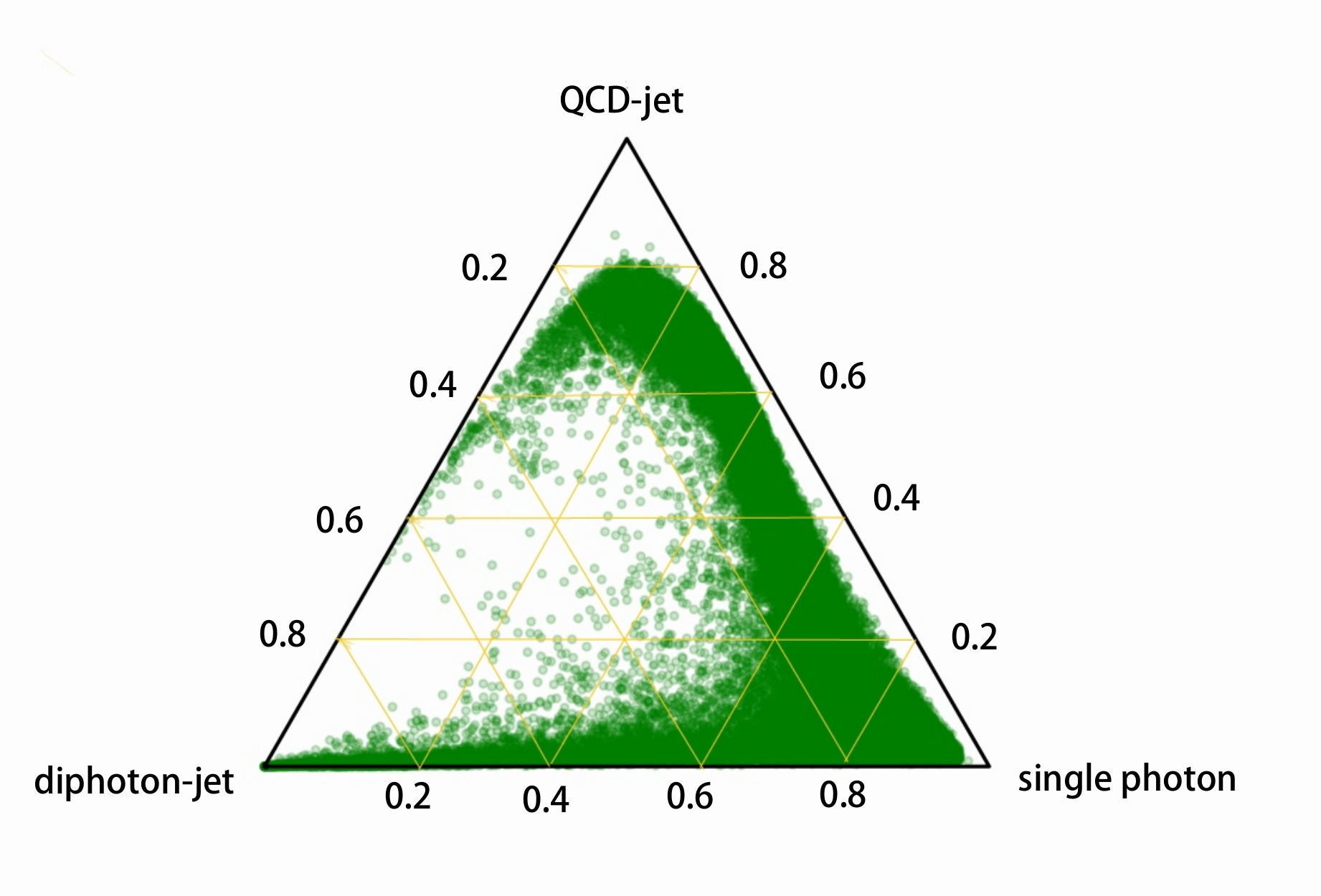}
\includegraphics[width=4.5cm]{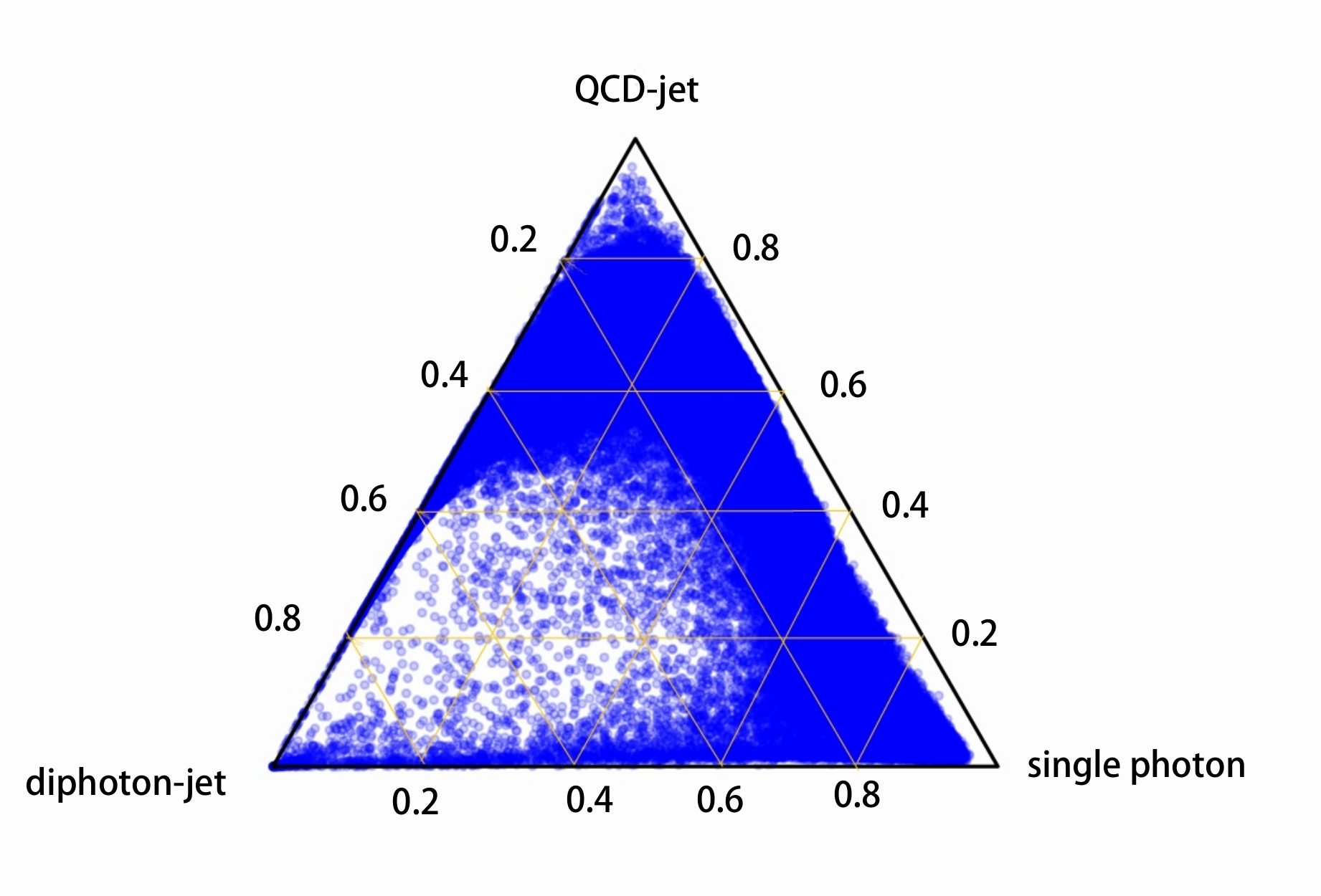}
\caption{A ternary plot for the diphoton-jet events, the single photon events and the QCD-jet events. The red, green and blue points represent the diphoton-jet events, the single photon events and the QCD-jet events, respectively.}
\label{ternary1}
\end{center}
\end{figure*}

\begin{figure*}[ht]
\begin{center}
\includegraphics[width=4.5cm]{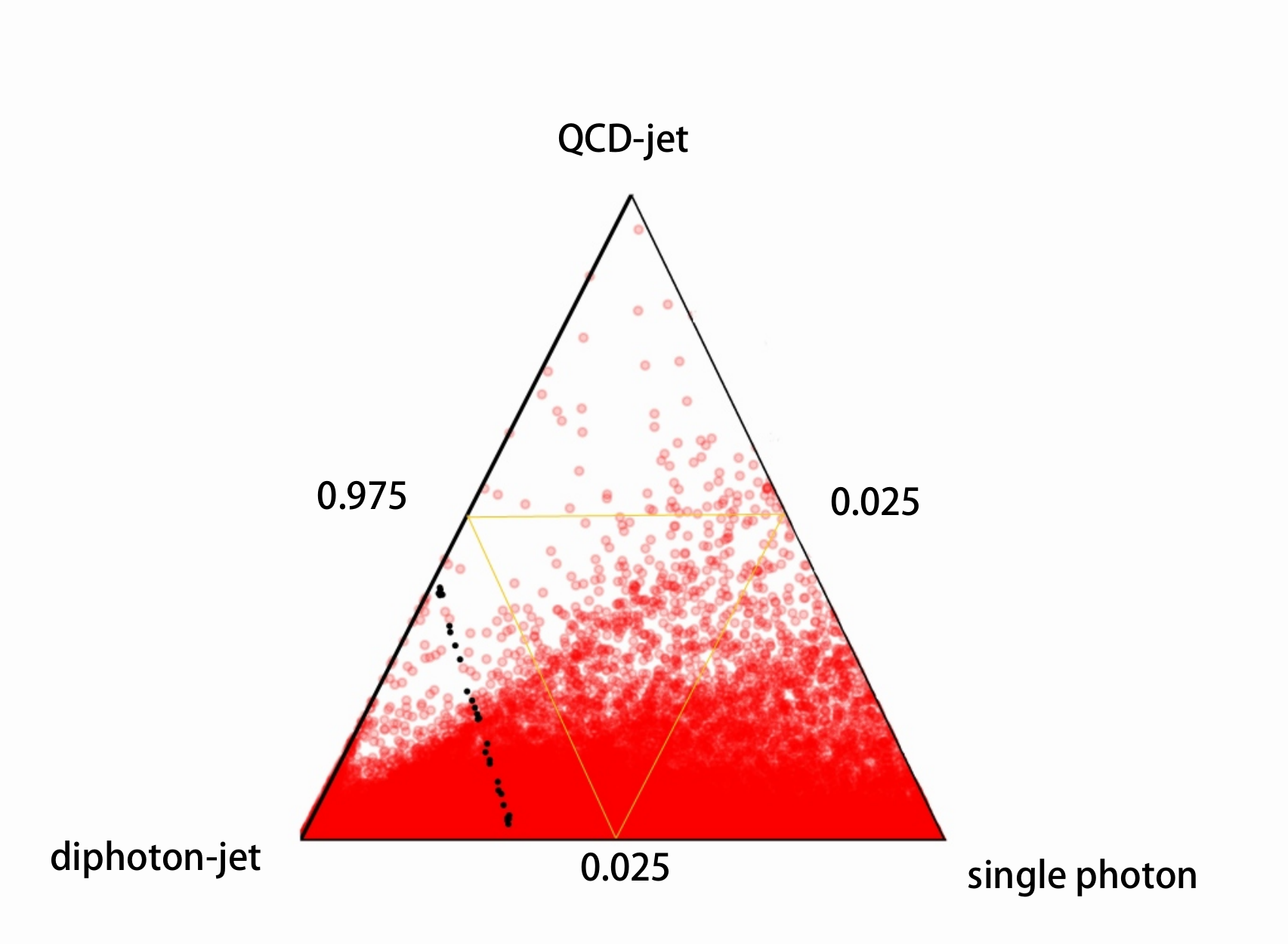}
\includegraphics[width=4.7cm]{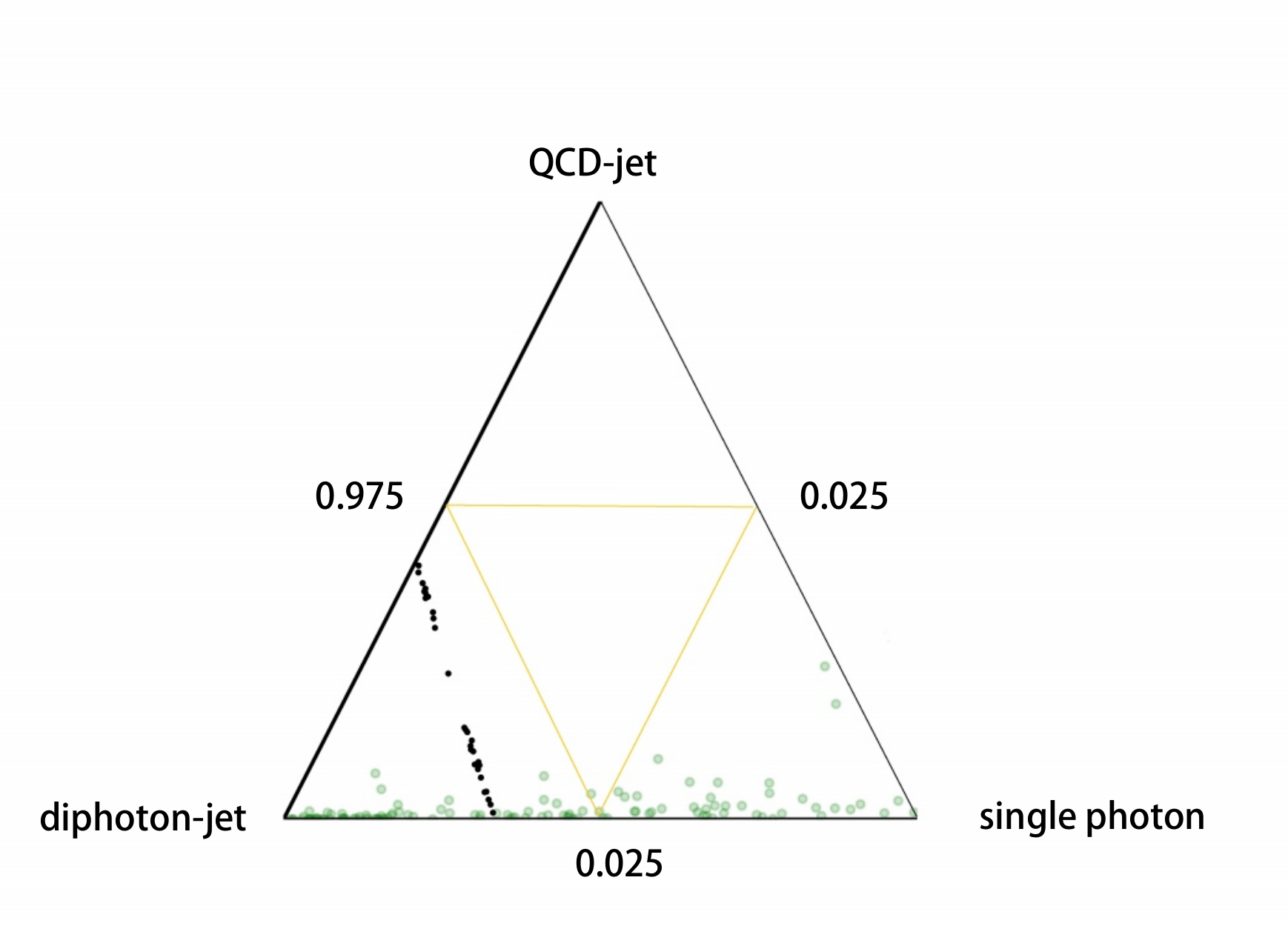}
\includegraphics[width=4cm]{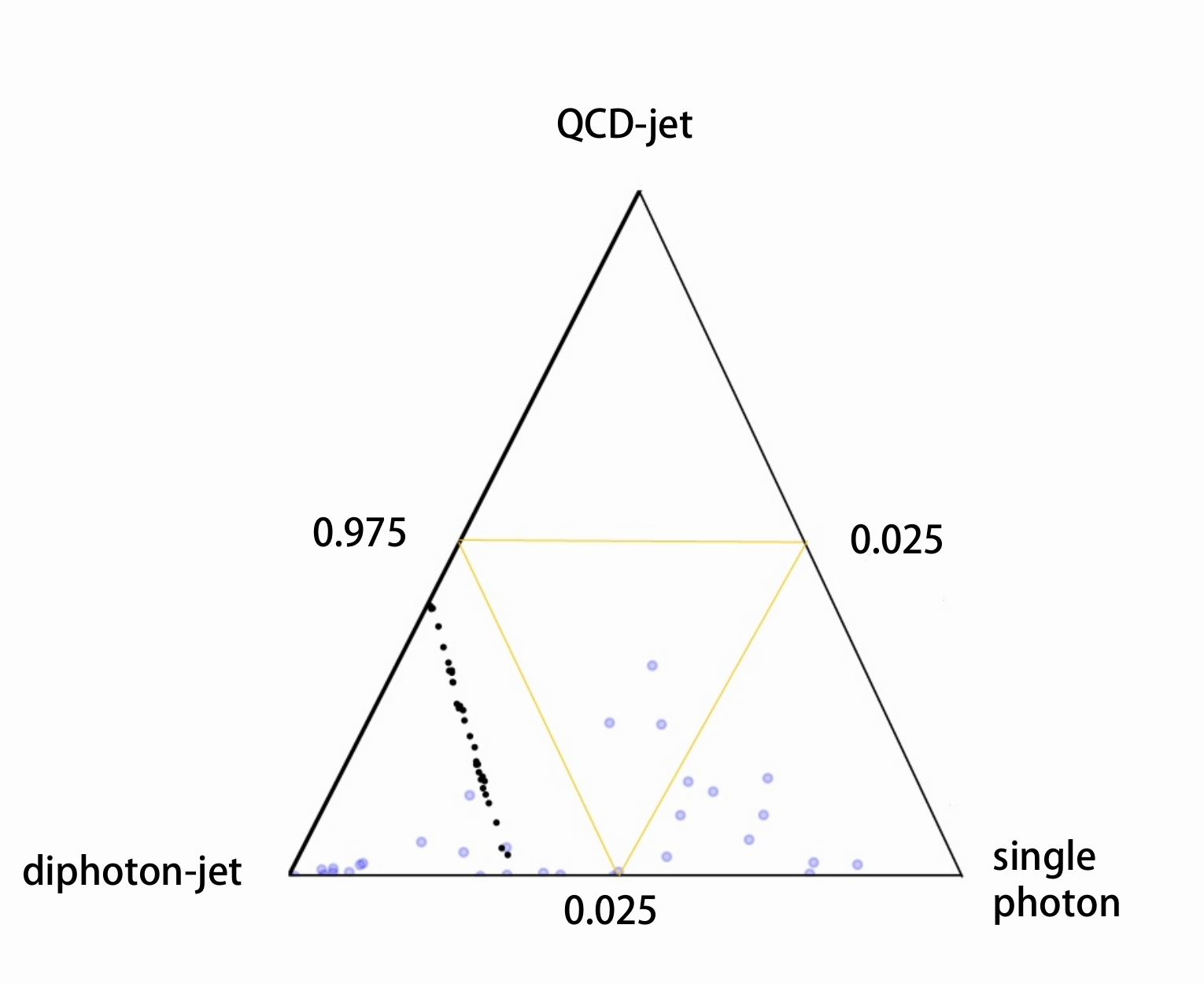}
\caption{The optimal cut for the ALP-strahlung production process $pp \to W^\pm a$ based on the selection NN with $g_{a\gamma\gamma}=0.64$ TeV$^{-1}$ and $m_{a}=3$ GeV. The red, green and blue points represent the diphoton-jet events, the single photon events and the QCD-jet events while the black line is the optimal cut.}
\label{ternary2}
\end{center}
\end{figure*}

After training, we ﬁnd the optimal model parameters and the jet-tagging CNN is tested on the jets in the remaining signal and background events based on the best model. The resulting jet tagging probabilities are fed into the jet selection NN to optimize the jet selection efficiency for the diphoton-jet, the single photon and the QCD-jet, respectively. We adopt the widely used ternary plot to present the jet tagging probabilities for each jet in the validation set. As shown in Fig.~\ref{ternary1}, the red, green and blue points denote the diphoton-jet, the single photon and the QCD-jet, respectively. Same as before, the benchmark point is chosen as $g_{a\gamma\gamma}=0.64$ TeV$^{-1}$ and $m_{a}=3$ GeV. It is clear that the size of the validation data set for the CNN is same as the size of the training data set for the jet selection NN.

In the following, we will employ a simple NN to obtain the optimal cut to maximize the statistic significance for $pp \to W^\pm a$ and $pp \to Za$ processes. The integrated luminosity is set to 3000 fb$^{-1}$ at the 14 TeV LHC. In Fig.~\ref{ternary2}, we present the optimal cut for the ALP-strahlung production process $pp \to W^\pm a$ based on the selection NN with $g_{a\gamma\gamma}=0.64$ TeV$^{-1}$ and $m_{a}=3$ GeV.

In order to estimate the statistical error, we train seven selection neural networks to obtain the mean detection significance and its standard deviation. For the diphoton-jet signal events and the single photon background events, we use seven different data sets, with each of them containing 300k events. The size of single photon background events which contains $W^\pm \gamma$ and $Z\gamma$ processes in the training data set are also 300k. The number of events from different processes are determined by the cross section. Note that, after jet selection cut, there are limited number of QCD-jets. However, generating QCD-jet events is computationally expensive. To ensure a stable statistics, we adopt the bootstrap sampling method for the QCD-jet background events which contains $jj$, $W^\pm j$, $t\bar{t}$, $tj$ and $Zj$ processes. At each time, 3M QCD-jets are randomly sampled from all the QCD-jet samples. The number of events from different processes are also determined by the cross sections. The event number in the training data set can ensure that there are at least 10 events left after any cut imposed by the neural network regardless of signal or backgrounds.
The optimal detection significance for the ALP-strahlung production processes $pp \to W^\pm a$ and $pp \to Za$ obtained from seven simple NNs are shown in Tables~\ref{table1} and \ref{table3}.

\begin{table}[ht]
\begin{center}\begin{tabular}{|c|c|c|c|c|c|c|c|c|c|c|}
\hline  & \tabincell{c}{1} &  \tabincell{c}{2} & \tabincell{c}{3} & \tabincell{c}{4}  &\tabincell{c}{5} & \tabincell{c}{6} & \tabincell{c}{7}   \\
\hline ALP Mass/GeV &\multicolumn{7}{c|}{Significance} \\ 
\hline  \tabincell{c}{0.3} & 0.12 & 0.16 & 0.21 & 0.15 & 0.13 & 0.22 & 0.18 \\
\hline  \tabincell{c}{0.5} & 0.31 & 0.39 & 0.41 & 0.33 & 0.26 & 0.29 & 0.32 \\
\hline  \tabincell{c}{1} & 0.54 & 0.51 & 0.45 & 0.49 & 0.44 & 0.50 & 0.44 \\
\hline  \tabincell{c}{2} & 0.73 & 0.86 & 0.81 & 0.69 & 0.66 & 0.61 & 0.69 \\
\hline  \tabincell{c}{3} & 1.49 & 1.52 & 1.47 & 1.36 & 1.41 & 1.39 & 1.44 \\
\hline  \tabincell{c}{4} & 1.99 & 2.05 & 2.11 & 2.07 & 2.14 & 1.95 & 1.98 \\
\hline  \tabincell{c}{5} & 2.86 & 2.65 & 2.65 & 2.74 & 2.88 & 2.71 & 2.63 \\
\hline  \tabincell{c}{6} & 2.86 & 2.25 & 2.35 & 2.54 & 2.48 & 2.51 & 2.43 \\
\hline  \tabincell{c}{7} & 2.71 & 2.52 & 2.42 & 3.08 & 2.63 & 2.99 & 2.80\\
\hline  \tabincell{c}{8} & 2.02 & 1.99 & 2.05 & 2.20 & 2.17 & 2.06 & 2.21 \\
\hline  \tabincell{c}{9} & 2.07 & 2.01 & 2.17 & 2.10 & 2.16 & 2.04 & 2.07 \\
\hline  \tabincell{c}{10} & 1.44 & 1.32 & 1.28 & 1.36 & 1.49 & 1.39 & 1.46 \\
\hline \end{tabular} \caption{The optimal significance for the ALP-strahlung production process $pp \to W^\pm a$ obtained from the trainings of seven significance neural networks. The benchmark point is chosen as $g_{a\gamma\gamma}=0.64$ TeV$^{-1}$. }
\label{table1}
\end{center}
\end{table}

\begin{table}[ht]
\begin{center}\begin{tabular}{|c|c|c|c|c|c|c|c|c|c|c|}
\hline   & \tabincell{c}{1} &  \tabincell{c}{2} & \tabincell{c}{3} & \tabincell{c}{4}  &\tabincell{c}{5} & \tabincell{c}{6} & \tabincell{c}{7}  \\
\hline ALP Mass/GeV &\multicolumn{7}{c|}{Significance} \\ 
\hline  \tabincell{c}{0.3} & 0.29 & 0.21 & 0.18 & 0.24 & 0.27 & 0.22 & 0.32\\
\hline  \tabincell{c}{0.5} & 0.77 & 0.72 & 0.83 & 0.69 & 0.71 & 0.66 & 0.64 \\
\hline  \tabincell{c}{1} & 0.98 & 0.91 & 0.87 & 0.85 & 0.93 & 0.88 & 0.95 \\
\hline  \tabincell{c}{2} & 1.89 & 1.91 & 1.85 & 1.93 & 1.90 & 1.82 & 1.88 \\
\hline  \tabincell{c}{3} & 2.08 & 2.15 & 2.11 & 2.17 & 2.04 & 2.12 & 2.06 \\
\hline  \tabincell{c}{4} & 2.49 & 2.32 & 2.38 & 2.63 & 2.31 & 2.30 & 2.33 \\
\hline  \tabincell{c}{5} & 2.06 & 1.98 & 1.93 & 2.15 & 1.99 & 2.14 & 2.02 \\
\hline  \tabincell{c}{6} & 1.75 & 1.70 & 1.68 & 1.72 & 1.76 & 1.69 & 1.73 \\
\hline  \tabincell{c}{7} & 1.66 & 1.45 & 1.47 & 1.70 & 1.50 & 1.55 & 1.44 \\
\hline  \tabincell{c}{8} & 1.61 & 1.51 & 1.49 & 1.53 & 1.49 & 1.68 & 1.66 \\
\hline  \tabincell{c}{9} & 1.44 & 1.64 & 1.52 & 1.49 & 1.55 & 1.59 & 1.65 \\
\hline  \tabincell{c}{10} & 1.41 & 1.56 & 1.59 & 1.45 & 1.52 & 1.38 & 1.36 \\
\hline \end{tabular} \caption{The optimal significance for the ALP-strahlung production process $pp \to Za$ obtained from the trainings of seven significance neural networks. The benchmark point is chosen as $g_{a\gamma\gamma}=0.64$ TeV$^{-1}$. }
\label{table3}
\end{center}
\end{table}

\begin{figure}[ht]
    \centering
    \includegraphics[width=12cm]{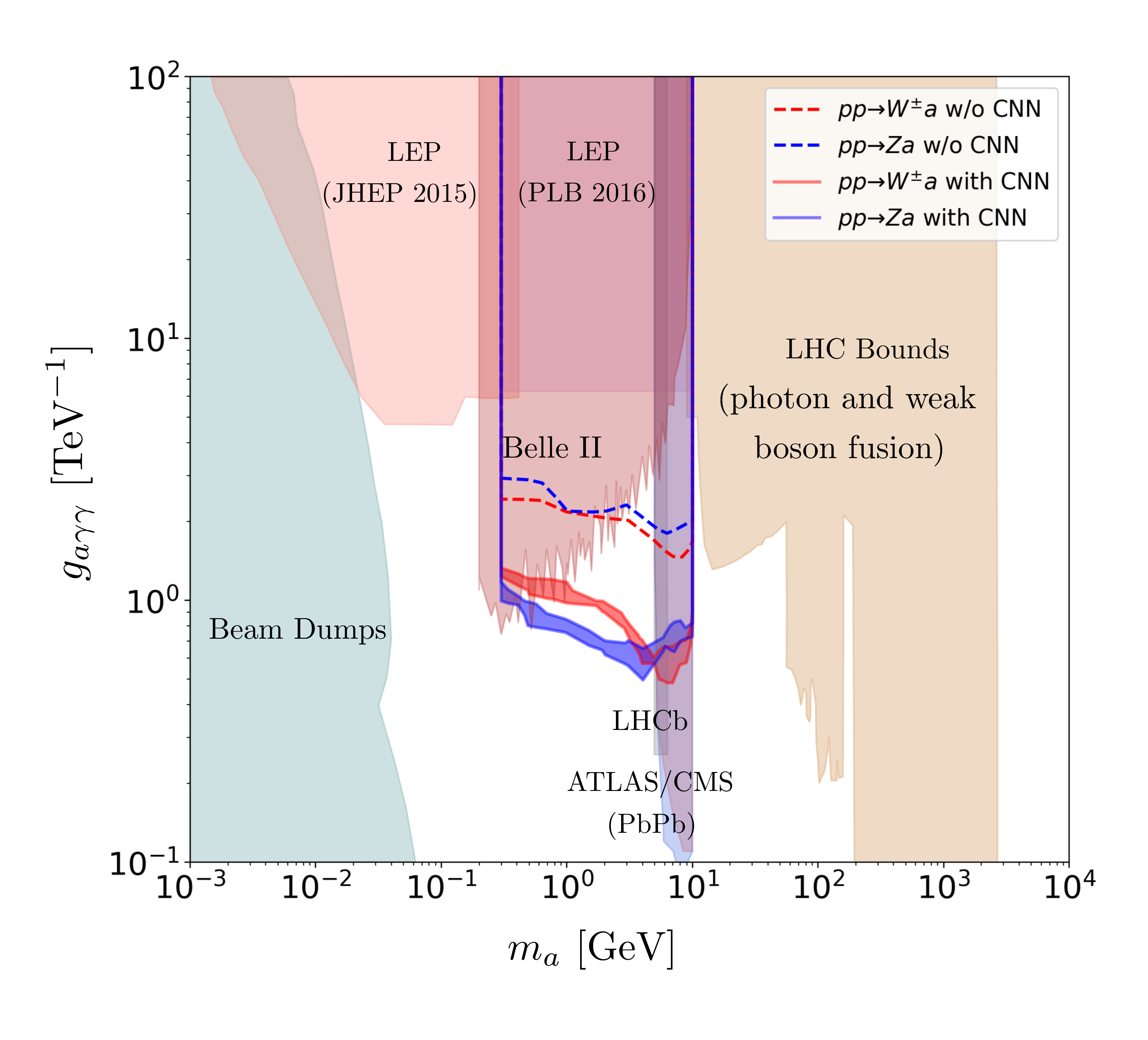}
    \caption{The $2\sigma$ bounds on the ALP-photon coupling $g_{a\gamma\gamma}$ versus $m_{a}$ plane. 
    Regions above the red and blue dashed lines are excluded by $pp \to W^\pm/Z a$ at the 14 TeV LHC with an integrated luminosity 3000 fb$^{-1}$, without using CNN~\cite{Wang:2021uyb}. Regions above red and blue bands are excluded by this work using CNN. Other bounds shown are from LEP~\cite{Mimasu:2014nea,Jaeckel:2015jla}, L3~\cite{Adriani:1992zm}, Belle II~\cite{Abudin_n_2020}, LHCb~\cite{CidVidal:2018blh}, ATLAS/CMS~\cite{Knapen:2016moh,Jaeckel:2015jla,ATLAS:2012fgo,Jaeckel:2012yz}, ATLAS/CMS(PbPb)~\cite{denterria2021collider} and Beam Dumps~\cite{Dobrich:2015jyk}.}
    \label{cm}
\end{figure}

In Fig.~\Ref{cm}, we present the $2\sigma$ bounds on the ALP-photon coupling $g_{a\gamma\gamma}$ versus  $m_{a}$ from the ALP-strahlung production processes $pp \to W^\pm/Z a$. Based on the results of the seven simple NNs, we calculate the mean significance and the corresponding standard deviation for all benchmark points and we draw two exclusion bands where each band has a width of 10 standard deviations. The red band indicates the $2\sigma$ bounds for $pp \to W^\pm a$ process while the blue band indicates the $2\sigma$ bounds for $pp \to Za$ process. As mentioned in Section II, the diphoton-jet we defined in this paper can only be applied to the ALP mass range from a few hundred MeV to 10 GeV. When the ALP mass is larger than 10 GeV, the ALP will decay to two separated photons and are detected as 2$\gamma$ events. When the ALP mass is less than a few hundred MeV, the ALP will be highly boosted and decay into a pair of highly collimated photons. The reconstruction of such a low mass resonance in our phenomenological study is very challenging so that we focus on the ALP mass range of 0.3 GeV to 10 GeV.


For comparison, we also present the previous constraints such as the beam dump searches for short-lived axions, the LEP searches for ALPs via $e^+ e^- \to 2\gamma/3\gamma$ processes, the di-photon resonance around the $B_{s}$ mass between 4.9 GeV and 6.3 GeV from the LHCb, the isolated and energetic photons produced by the hadronic decay of $Z$ boson at the L3. Since the resonant searches based on Upsilon meson decay process $\Upsilon \to a\gamma$ at Babar can only be used to scenarios with a non-zero ALP-gluon coupling while in our analysis we turn off the ALP-gluon coupling, we do not include the BaBar limits. 
Note that the searches for $\gamma\gamma$ resonances in photon and weak boson fusion processes by ATLAS and CMS~\cite{Knapen:2016moh,Jaeckel:2015jla,ATLAS:2012fgo,Jaeckel:2012yz} are sensitive only when $m_{a}>$10 GeV. Recently, The $\gamma\gamma \to a \to \gamma\gamma$ 
searches have been utilized in electromagnetic PbPb collisions at the LHC by CMS and ATLAS~\cite{CMS:2018erd,ATLAS:2020hii}, which provide the current most competitive ALPs limits in the mass range of 5 GeV $<m_a<$ 100 GeV, as shown in Fig.~\Ref{cm}. Besides, $e^{+}e^{-}\to\gamma a, a \to \gamma\gamma$ processes have been searched in the mass range 0.2 GeV $<m_a<$ 9.7 GeV in the Belle II~\cite{Abudin_n_2020}, as shown in Fig.~\Ref{cm}.
Moreover, we also show the constraints from Ref.~\cite{Wang:2021uyb} , in which the electroweak ALP is probed via the ALP-strahlung production processes $pp \to W^{\pm}/Z a$ in the mass range of 0.3 GeV $<m_a<$ 10 GeV at the 14 TeV HL-LHC based on jet substructure variables and BDT method. The constraints are depicted as red dashed line and blue dashed line to present the $2\sigma$ bound for the $pp \to W^\pm a$ process and the $pp \to Za$ process, respectively. Since the results in Ref.~\cite{Wang:2021uyb} is based on conventional jet substructure variables and BDT, we tagged these two dashed lines with ``pp$\rightarrow{W^\pm a}$ without CNN'' and ``pp$\rightarrow{Za}$ without CNN''. As shown in Fig.~\Ref{cm}, the shapes of the red and blue bands in this work based on CNN are similar to the shapes of the dashed red and blue lines in Ref.~\cite{Wang:2021uyb} based on jet substructure variables which could cover part of the triangle region between the Belle II bound and the ATLAS/CMS(PbPb) bound. The exclusion limits are weaker when $m_{a}$ is close to 0.3 GeV and 10 GeV since the diphoton-jet signal feature is not clear with either being detected as 2$\gamma$ events or being tagged as a single photon in our analysis. The best exclusion limits from the processes $pp \to W^\pm a$ and $pp \to Z a$ are obtained when $m_{a}$ are in the range of 5 to 7 GeV and 3 to 5 GeV, respectively. For the same ALP-strahlung process, diphoton-jet tagging based on CNN in this work is much better than that based on jet substructure variables analysis in Ref.~\cite{Wang:2021uyb}.

We find that our approach based on the jet-tagging CNN and the detection significance optimization NN can improve the current LHC sensitivities to the ALP mass from 5 GeV to 0.3 GeV in the case of vanishing ALP-gluon coupling. Moreover, it can greatly surpass the existing LEP bounds, Belle II constraints and the limits from Ref.~\cite{Wang:2021uyb}. For instance, $g_{a\gamma\gamma}$ in our study can be excluded to 1.1 TeV$^{-1}$ at $m_a=0.3$ GeV and 0.5 TeV$^{-1}$ at $m_a=5$ GeV. It should be noted that our obtained limit from the HL-LHC is stronger than that from current Belle-II data. However, the future Belle II with full integrated luminosity of 50 ab$^{-1}$ may provide more stringent constraints than ours~\cite{Dolan:2017osp}.

In this work, we mainly investigate the potential of utilizing the CNN method to distinguish the ALP diphoton-jet events from the single photon and QCD jets events. But it is worth to note that additional standard kinematic cuts exploiting the full kinematical properties of the signal and background events could be used to further suppress the backgrounds and enhance the sensitivity. For example, we note that the resulting QCD-dijet events is the dominant background for the signal process $ pp \to W^\pm a$. Therefore, isolating photon and a cut on the missing transverse energy would be helpful to reduce the such a background in our analysis of the process $ pp \to W^\pm a$. Our results obtained in this paper should be considered as a lower limit on top of which further improvements can be implemented. Besides, we should mention that our sensitivity analysis would be affected by the systematic uncertainties, such as the calibration of the jet energy scale. However, this will need a full simulation of detector and real data. Since the realistic detector performances of the HL-LHC are not still available, we do not include the systematic uncertainties in the current calculation. An updated analysis will be done in our future work.

\section{Conclusions}

In this work we studied the ALP-strahlung production processes $pp \to W^\pm a, Z a$ in the mass range of 0.3 GeV $<m_a<$ 10 GeV at the 14 TeV HL-LHC. Since the two photons from the ALP decay are highly collimated for such a light ALP, we designed a jet-tagging CNN to discriminate our signal from the QCD-jets and the single photon backgrounds based on the jet-image notion and proposed a detection significance optimization NN to search for the optimal cut to maximize the statistic significance for the ALP-strahlung production processes $pp \to W^\pm a, Z a$. With the help of machine learning techniques, we obtained the $2\sigma$ bounds on the ALP-photon coupling $g_{a\gamma\gamma}$ versus the ALP mass $m_{a}$. The coupling $g_{a\gamma\gamma}>$ 1.1 TeV$^{-1}$ at $m_a=0.3$ GeV and $g_{a\gamma\gamma}>$ 0.5 TeV$^{-1}$ at $m_a=5$ GeV can be excluded at $2\sigma$ level at the 14 TeV LHC with an integrated luminosity of 3000 fb$^{-1}$. This shows that our approach can extend the current LHC bounds on the ALP mass from 5 GeV to 0.3 GeV and the obtained bounds are stronger than the existing other limits. 

\acknowledgments
This work was supported by the National Natural Science Foundation of China 
(NNSFC) under grant Nos. 11821505, 12075300, 11947118,
by Peng-Huan-Wu Theoretical Physics Innovation Center (12047503),
by the CAS Center for Excellence in Particle Physics (CCEPP), 
by the CAS Key Research Program of Frontier Sciences, 
by a Key R\&D Program of Ministry of Science and Technology of China
under number 2017YFA0402204,
and by the Key Research Program of the Chinese Academy of Sciences, Grant NO. XDPB15. 

\bibliography{ref}
\end{document}